\documentclass[useAMS,usenatbib]{mnras}
\usepackage{graphicx}
\usepackage{caption}
\usepackage{subcaption}
\usepackage{amsmath}
\usepackage{array}
\usepackage[T1]{fontenc}
\usepackage{threeparttable}
\def\be{\begin{equation}}
\def\ee{\end{equation}}
\usepackage{float}
\usepackage{amssymb}
\usepackage[usenames,dvipsnames,svgnames,Table]{xcolor}
\usepackage{enumitem}

\usepackage{mwe}
\usepackage{hyperref}

\title[An expanding corona in multiple X-ray binaries]{Evidence for an expanding corona based on spectral-timing modelling of multiple black hole X-ray binaries}

\author[Cao et al.]
    {Zheng Cao$^{1,\,2,\,3}$\thanks{E-mail: z.cao@sron.nl}, Matteo Lucchini$^{1,\,4}$, Sera Markoff$^{1,\,5}$, Riley M. T. Connors$^{6}$, and \newauthor{Victoria Grinberg$^{7}$}
       \\
     $^1$Anton Pannekoek Institute for Astronomy, University of Amsterdam, Science Park 904, 1098 XH, Amsterdam, The Netherlands\\
     $^2$SRON, Netherlands Institute for Space Research, Sorbonnelaan 2, 3584 CA, Utrecht, The Netherlands\\
     $^3$Department of Astrophysics/IMAPP, Radboud University, P.O. Box 9010, 6500 GL, Nijmegen, The Netherlands\\
     $^4$MIT Kavli Institute for Astrophysics and Space Research, MIT, 70 Vassar Street, Cambridge, MA 02139, USA\\
     $^5$Gravitation Astroparticle Physics Amsterdam (GRAPPA) Institute, University of Amsterdam, Science Park 904,\\
     1098 XH Amsterdam, The Netherlands\\
     $^{6}$Cahill Center for Astronomy and Astrophysics, California Institute of Technology, Pasadena, CA 91125, USA\\
     $^{7}$European Space Agency (ESA), European Space Research and Technology Centre (ESTEC), Keplerlaan 1, \\
     2201 AZ Noordwijk, The Netherlands\\
     }
     \date{}

\pagerange{\pageref{firstpage}--\pageref{lastpage}} \pubyear{xxx}

\begin{document}

\label{firstpage}

\maketitle

%%%%%%%%%%%%%%%%%%%%%%%%%%%%%%%%%%%%
%%%%%%% ABSTRACT %%%%%%%%%%%%%%%%%%%%%%
\begin{abstract}

Galactic black hole X-ray binaries (BHXBs) provide excellent laboratories to study accretion, as their relatively quick evolution allows us to monitor large changes in the in-flowing and/or out-flowing material over human timescales. However, the details of how the inflow-outflow coupling evolves during a BHXB outburst remain an area of active debate. In this work we attempt to probe the physical changes underlying the system evolution, by performing a systematic analysis of the multi-wavelength data of three BHXB sources: XTE J1752-223, MAXI J1659-152, and XTE J1650-500, during hard and hard-intermediate states. Using the power spectral hue which characterises the X-ray variability properties, we identify several clusters of BHXB epochs and perform the joint multi-wavelength spectral modelling to test their commonality with a physical jet model. Under the assumption that the corona is related to the base of the jet, we find that the power spectral hue traces the variation of the coronal radius (from ${\sim}10R_{g}-{\sim}40R_{g}$) in multiple BHXBs at hard and hard-intermediate states, and that the data are consistent with moderately truncated accretion discs ($<25R_{g}$) during hard-intermediate states. We also find that all epochs of low disc reflection have high hue located near the hard-intermediate to soft-intermediate state transition, indicating that in these states the vertical extent of the corona and/or its bulk speed are increasing. Our results link the geometrical similarity in the corona among multiple BHXB sources to their timing characteristics, and probe the corona responding to the disc-jet interactions at hard and intermediate states during outbursts.

\end{abstract}


%%%% INTRODUCTION %%%%%%%%%%%%%%%%%%%%
\section{Introduction}
\label{sec:intro}

Galactic low mass black hole X-ray binaries (BHXBs) are binary systems with a stellar-mass black hole accreting mass from its low mass companion star. The majority of Galactic BHXBs are transient X-ray sources that go from quiescence to outburst phases \citep[e.g.][]{elvis1975discovery,tanaka1996x,chen1997properties,homan2005evolution,remillard2006x,corral2016blackcat,tetarenko2016watchdog}. The outbursts can last weeks to months, usually exhibiting transitions between a well-defined series of X-ray spectral states \citep[e.g.][]{nowak1995toward,esin1997advection,mendez1998canonical,homan2005evolution,remillard2006x,belloni2010states}.

During a typical outburst, the X-ray spectrum of a source is initially at the hard state (HS), dominated by a hard power-law component as the luminosity rises from quiescence to highly luminous states. This power-law component is thought to be generated due to the inverse-Comptonisation of soft-disc photons by hot electrons in a loosely defined "compact corona". Different proposed models for this corona include the innermost regions of a radiatively inefficient accretion flow (RIAF, e.g. \citealt{narayan1994advection,esin1997advection,yuan2007modeling,veledina2013hot}), an extended slab corona over the disc \citep[e.g.][]{haardt1993x,haardt1994model}, or a "lamp-post", compact magnetised region above the disc and/or moving vertically above the black hole \citep[e.g.][]{matt1992iron,martocchia1996iron,beloborodov1999electron,merloni2002coronal}. A physical realisation of this model would be if the lamp-post is associated with the base of the jet \citep[e.g.][]{markoff2004constraining,markoff2005going,maitra2009constraining,dauser2013irradiation,miller2015new,furst2015complex,lucchini2021}. As the accretion rate increases, the X-ray spectrum of the source softens, becoming increasingly dominated by the thermal emission of a geometrically thin, radiatively efficient accretion disc \citep{shakura1973black}, and the hard power-law component softens and fades. This is the so-called "soft state" (SS). Eventually, the source will again decrease in luminosity and transition back to the HS before returning to quiescence. The transitional spectral states are referred to as the hard/soft intermediate states (HIMS/SIMS) \citep[][]{homan2005evolution,belloni2010states}. Roughly 40\% of outbursts fail to reach the SS; these are termed "failed" outbursts \citep{tetarenko2016watchdog}.

Compact, magnetised jets are one of two types of characteristic outflows associated with BHXB outbursts. Jets are collimated relativistic streams of plasma launched from the inner accretion region. Particles are also accelerated within BHXB jets, similarly to AGN \citep{markoff2001jet,fender2005energization,romero2005gamma,bosch2006broadband,kylafis2012formation,malzac2013internal,kantzas2021new}, leading to synchrotron emission by relativistic electrons. The stratification of optically-thick regions generates a flat/inverted radio-through-IR spectrum, which is one of the hallmarks of the presence of jets \citep{blandford1979relativistic,hjellming1988radio,fender2000very,fender2004towards,corbel2000coupling}. During a typical BHXB outburst, compact jets are present during the HS, and suppressed during the SS/quenched before the SS (\citealt{tananbaum1972observation,fender1999quenching,fender2004towards,gallo2003universal,belloni2010states}, but see \citealt{drappeau2017dark}).

Although the underlying physics is not yet understood, the coupling between inflowing and outflowing material results in a correlation between the radio and X-ray luminosity \citep[e.g][]{hannikainen1998most,corbel2000coupling,corbel2003radio,corbel2013universal,gallo2003universal,gallo2014radio}. This correlation can be extended to accreting supermassive black holes in jetted active galactic nuclei (AGN) by adding a term accounting for the black hole mass \citep{merloni2003fundamental,falcke2004scheme,plotkin2012using}; this extension is referred to as "the fundamental plane of black hole accretion". This connection between radio and X-ray emission, together with other observational similarities in such as the X-ray variability or the broadband spectral energy distributions \citep[e.g.][]{mchardy2006active,markoff2008results,markoff2015above}, supports the idea that (at least to first order) the accretion physics can be scaled between black holes across the mass scale. However, the viscous timescale in AGN is on the order of $\approx 10^{3-4}$ years; therefore, we can only rely on BHXBs to probe the jet-disc coupling of the same source at vastly different accretion rates. 

The X-ray light curves of BHXBs show variability across a wide range of timescales. The information carried by this short-term variability can be used to complement spectral data to characterise the accretion flow, typically by employing Fourier techniques \citep[e.g.][]{van1989fourier,wijnands1999broadband,klein2008identification,uttley2014x}. The continuum shape of the power spectral density (hereafter PSD) is strongly correlated with the X-ray spectral states, possibly because both are driven by the geometry of the system \citep[e.g.][]{done2006disc,done2007modelling,belloni2010states,ingram2012modelling,rapisarda2017modelling,ingram2019review}. Understanding both the spectral and timing properties of accreting black holes and how they are related to each other is very important since they are simultaneously produced by the same accretion process and together can provide additional independent constraints to models of black hole accretion.

\begin{figure}
    \centering
    \includegraphics[width=\linewidth]{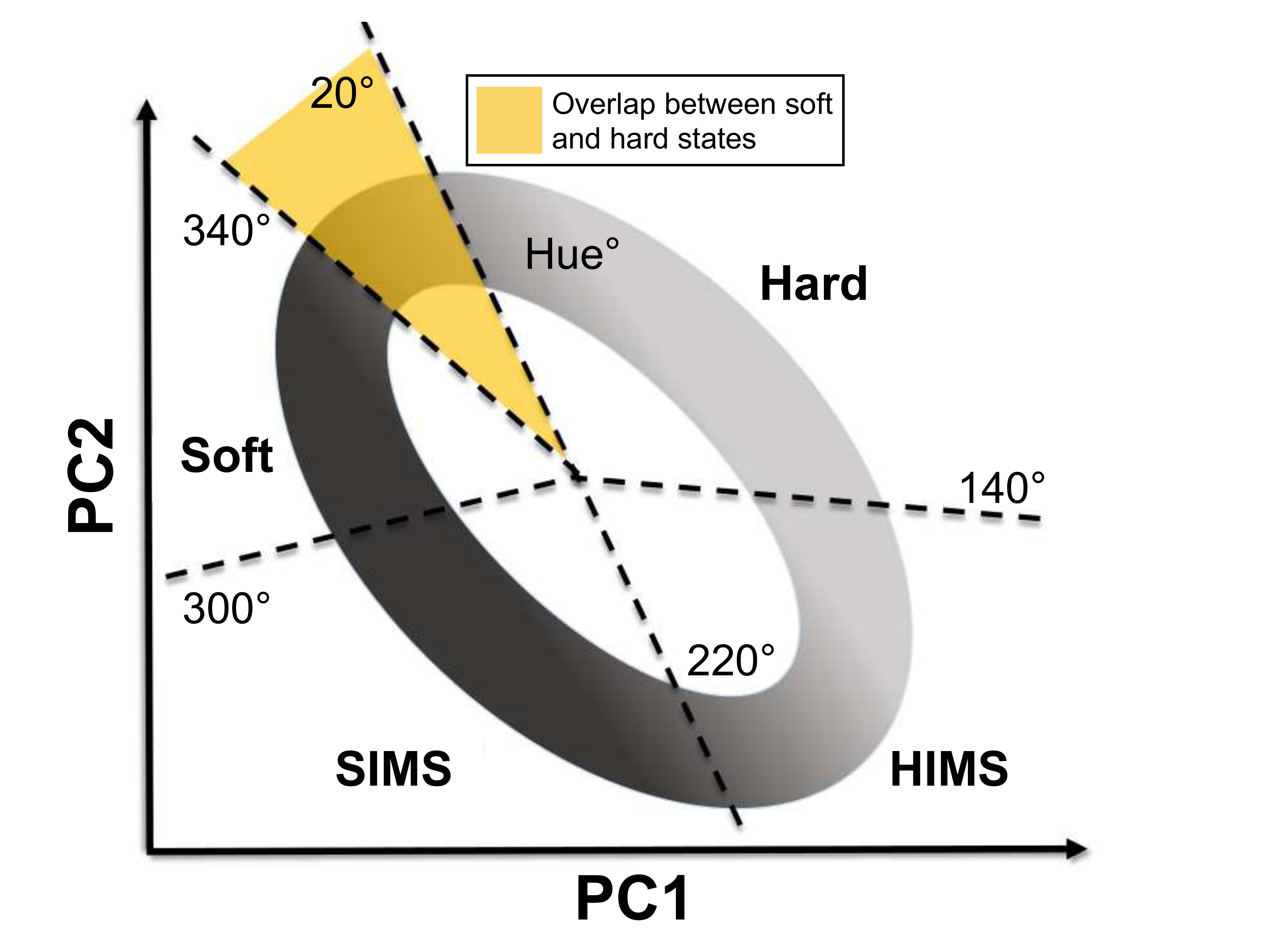}
    \caption{A conceptual illustration of the power-colour hue diagram, taken from \protect\cite{connors2019combining} based upon \protect\cite{heil2015power}. Both axes are in log scale. Hue is defined as the angular position on the diagram where 0$^{\circ}$ corresponds to the semi-major axis at 45$^{\circ}$ to the x- and y-axes. Dotted lines mark the regions typically corresponding to spectral states: soft, hard, hard intermediate state(HIMS), and soft intermediate state (SIMS); soft and hard states overlap in the top left of the diagram due to similar shapes in power density spectra, which can be distinguished only by spectral analysis. A typical BHXB outburst will start in quiescence, a low luminosity extension of the hard state, and evolve clockwise to the soft state, and then move anti-clockwise back to the hard states \citep{heil2015power}.}
    \label{fig:pc}
\end{figure}

\cite{heil2015power} introduced the new concept of power spectral "power-colours", defined as the ratio of the variances between two different Fourier frequency bands in the PSD, characterising the shape of the PSD independently of its normalisation (power colour index 1 = variance in 0.25-2.0 Hz/0.0039-0.031 Hz, and power colour index 2 = variance in 0.031-0.25 Hz/2.0-16.0 Hz, hereafter PC1 and PC2). By performing the analysis of 12 different BHXB sources, \cite{heil2015power} found that power colour provides a complementary phase-space to the spectrally-based hardness-intensity diagram in order to trace the evolution of BHXB outbursts via their angular positions on the power-colour diagram (PC1 vs. PC2), defined as the "hue" angles (illustrated in Fig.~\ref{fig:pc}). The correlation between the spectral hardness and the hue is shared among different sources, but the physical reason behind this empirical correlation is largely unknown. In this way, a typical outburst will start from the HS with a small hue angle (at the top left of the PC1/PC2 diagram), evolve clockwise through the HIMS/SIMS to the SS with a large hue on the power-colour diagram, and then return to quiescence following a similar counter-clockwise track.

In this paper, we continue the exploration begun in \cite{connors2019combining} and \cite{lucchini2021}, attempting to link the X-ray timing and multi-wavelength properties within the context of outflow-dominated BHXB models, including a more detailed coronal region. We perform a systematic study of multiple BHXB sources, combining the X-ray variability classification scheme of \cite{heil2015power} with broadband spectral energy distribution (SED) modelling during both hard states and hard-to-soft transitions. The goal of our modelling effort is to identify a consistent picture of the system evolution throughout outbursts. We do this by trying to identify possible trends in the geometry of the X-ray emitting region with respect to the spectral and timing evolution of the sources.

The paper is structured as follows. In Section \ref{sec:data} we describe the source selection and the data used in this work. In Section \ref{sec:model} we discuss our disc+jet model used in the spectral analysis, along with our assumptions. In Section \ref{sec:results} we present both the results of our broadband spectral fits and X-ray timing analysis, as well as the trends in the geometrical parameters we find in the long-term variability of our sources. In Section \ref{sec:discussion} we discuss our findings and in Section \ref{sec:conclusions} we summarise our results and conclude.

%%%% DATA %%%%%%%%%%%%%%%%%%%%
\section{Data}
\label{sec:data}

\subsection{Source Selection}

We define our sample as BHXB sources with more than one epoch of quasi-simultaneous (performed within a day of each other) radio, X-ray, and infrared/optical/UV observations available, with the goal of covering the broadband spectral evolution in both the jets and the accretion flow from HS to HIMS as accurately as possible. We also require that epochs should have power-law dominated X-ray spectra and a flat or inverted radio spectrum, as well as timing characteristics consistent with the HS or HIMS. We do not consider any SIMS/SS epochs as the compact jets are not active and the corona is sub-dominant. Because the power colours are only defined for the \textit{RXTE}/PCA energy band \citep{heil2015power}, we restrict our analysis to sources which have \textit{RXTE} (\citealt{jahoda1996orbit}) observations. Therefore, we do not include a few more recent BHXB sources in our search \citep[e.g.][]{uttley2018nicer,armas2019multiwavelength,shang2019evolution,xu2020studying}. Calibrating the power colours across different instruments is an interesting problem to be considered in future works.  

The database used for selecting the sources in this work is the \textit{Whole-sky Alberta Time-resolved Comprehensive black-Hole Database Of the Galaxy}, or \textit{WATCHDOG} \citep{tetarenko2016watchdog}. This comprehensive database includes 47 transient and 10 persistent XRB sources with confirmed black holes or black hole candidates as the primary objects. The total number of transient outbursts recorded is more than 130 over the two decades considered in this database (1996-2015). We then searched the literature to check which of these sources fulfils our selection criteria. Three optimal sources were identified for this work: XTE J1752-223, MAXI J1659-152, and XTE J1650-500. The full details of the observations analysed can be found in the Appendix.

\subsection{Data reduction and collection}

We complement our \textit{RXTE} hard X-ray data with soft X-ray data from the XRT instrument on-board \textit{Neil Gehrels Swift} satellite, and optical or UV data from \textit{Swift}/UVOT. X-ray spectra from \textit{Swift}/XRT are extracted using the online \textit{xrtpipeline} provided by \cite{evans2009methods}\footnote{ https://www.swift.ac.uk/user\_objects/}; we only consider the $1.0-10\,\rm{keV}$ energy range in order to avoid detector features below $1\,\rm{keV}$. Spectra are grouped to reach a minimum signal-to-noise ratio of 20 before spectral analysis. We include any available \textit{Swift}/UVOT observations taken together with the \textit{Swift}/XRT data. All the optical/UV images of \textit{Swift}/UVOT, if available, are taken directly from the archive, and \textit{uvotsource} is used to extract the photometry of selected sources from the image. The source extraction regions are chosen to be circular, with the centres the same as the coordinates of our sources provided in the SIMBAD astronomy database\footnote{http://simbad.u-strasbg.fr/simbad/}, using a radius of 5 arcsec as suggested by the Swift team\footnote{https://www.swift.ac.uk/analysis/uvot/mag.php}. The backgrounds are chosen to be a larger circular area without any other sources and close to our targets, in order to minimise the impact of background fluctuations.

Radio, IR, and other optical/UV data in this work are taken from the literature on each source (for XTE J1752-223: \cite{brocksopp2013xte}; MAXI J1659-152: \cite{van2013broad}; XTE J1650-500: \cite{corbel2004origin,curran2012disentangling}). We perform the de-reddening of IR/optical/UV data ourselves to make sure the observations are consistently treated. The hydrogen column density $N_H$ in cm$^{-2}$ related to the galactic extinction for each target is adopted from existing publications. We estimate the extinction using the $A_V\sim N_H$ relation in \cite{foight2016probing}: $N_H=(2.87\pm0.12)\times10^{21}A_V$. Finally, we re-scale the extinction $A_V$ to various IR/optical/UV bands $A_{\omega}$ using the online tool\footnote{http://www.dougwelch.org/Acurve.html} based on the extinction law studied in \cite{cardelli1989relationship}. The measured IR/optical/UV flux in mJy at wavelength $\omega$ band is de-reddened by a factor of $10^{0.4A_\omega}$. This allows us to use consistent extinction/absorption in the IR/optical/UV and the X-ray bands, i.e. the $N_H$ in the X-ray extinction model \texttt{TBabs} \citep{wilms2000absorption} during spectral fits.

%%%% MODEL %%%%%%%%%%%%%%%%%%%%
\section{Multi-wavelength Jet Model}
\label{sec:model}

\begin{table*}
\caption[List of parameters and descriptions in the jet model]{A full list of parameters and descriptions in the jet model used in this work. Parameters in bold font are left free during spectral fitting, while the others are frozen.}
%\centering
\begin{tabular}{p{3cm}p{10cm}}
\hline
Parameter & Description \\
\hline
$\boldsymbol{N_j}$ & Total power channeled into the jet base, in units of Eddington luminosity $L_{\rm Edd}$.           \\
$\boldsymbol{R_0}$ & Radius of the cylindrical jet base/corona in units of $\rm{R_g}$.        \\
$\boldsymbol{T_e}$ & Temperature of the thermal electrons injected in the corona in units of $\rm{keV}$.        \\
$\boldsymbol{f_{\rm pl}}$ & Free parameter used to model inverted, rather than flat ($f_{\rm pl}=0$), radio spectra. \\
$\boldsymbol{R_{\rm in}}$ & Radius of the inner edge of the thin accretion disc in units of $\rm{R_g}$.        \\
$\boldsymbol{L_{\rm disc}}$ & Luminosity of the thin disc in units of $L_{\rm Edd}$.        \\
\hline
$M_{\rm BH}$, $\theta_i$, $d$ & Mass, inclination angle, and distance of the black hole. We fix these parameters to reported values in the literature. See Table~\ref{tb:srcpars} in Results.\\
$z_{\rm acc}=2000\,\rm{R_g}$ & Distance along the jet where the jet bulk acceleration ends and particle acceleration initiates.         \\
$\epsilon_{pl}=0.1$ & The fraction of initially thermal particles accelerated into a power-law tail at $z_{\rm acc}$.         \\
$p=2.3$ & Slope of the power-law of the non-thermal particle distribution. \\
$f_{b}=0.1$ & Adiabatic cooling efficiency. It sets the location of the cooling break in the radiating particle distribution.\\
$\beta_p=0.0315$ & Standard plasma-$\beta$ parameter at the base of the jet. A given set of $\gamma_{\rm acc}$ and $\sigma_{\rm f}$ determines the number ratio between protons and electron-positron pairs. Our choice corresponds to roughly $\sim 30$ pairs per proton. \\
$f_{\rm sc}=0.1$ & Efficiency parameter for particle acceleration, which sets the maximum energy cutoff of the particle power-law distribution. \\
$\sigma_{\rm f}=0.1$ & The final magnetisation parameter after the jet bulk acceleration stops at $z_{\rm acc}$. \\
$\gamma_{\rm acc}=3$ & The final Lorentz factor of the jet after the bulk acceleration. \\
\hline
\end{tabular}
\label{tb:pars}
\end{table*}

\begin{figure}
    \centering
    \includegraphics[width=\linewidth]{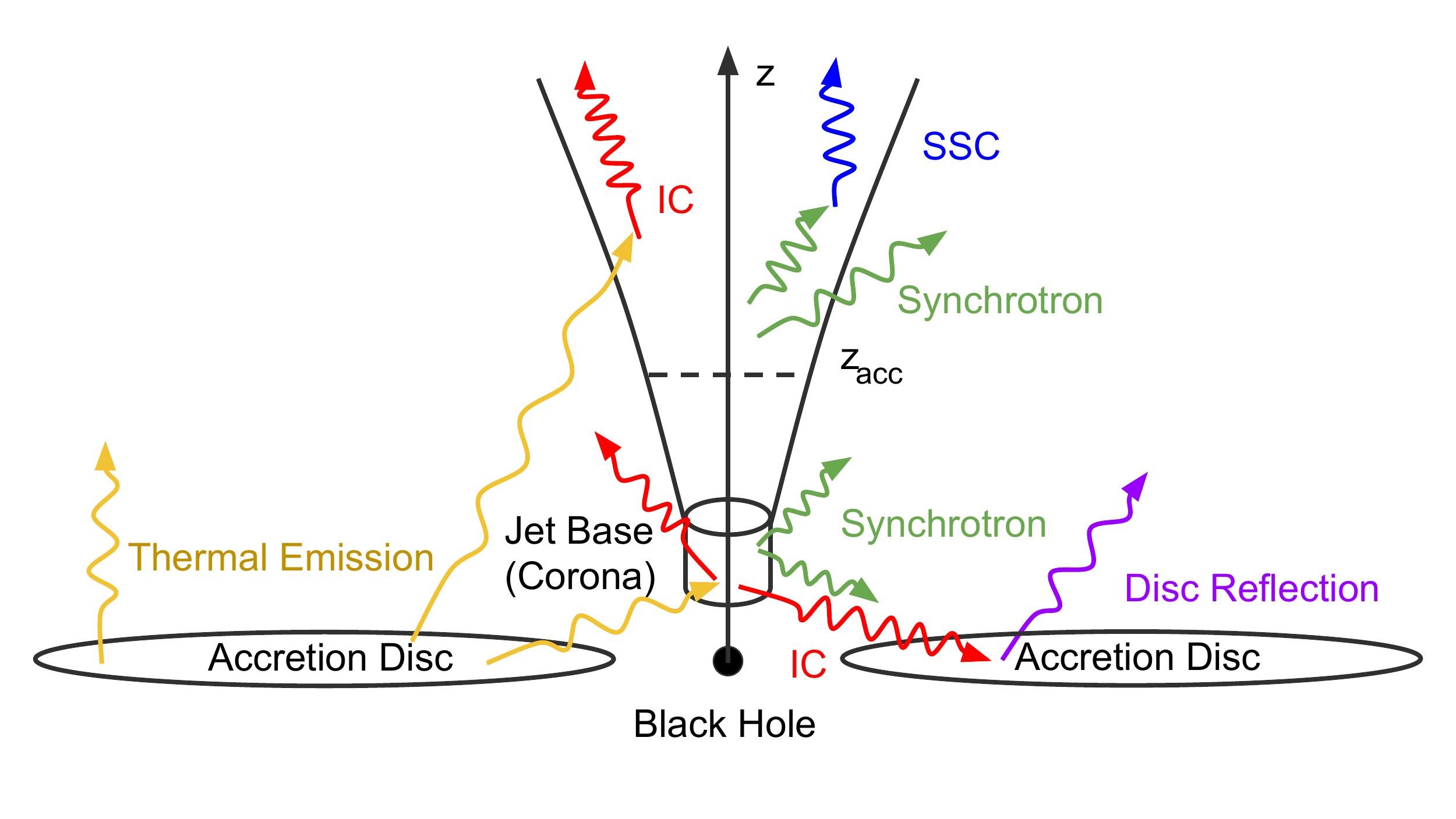};
\caption{Schematic of the emission model used in this work. The thermal photons from the accretion disc are inverse-Compton (IC) up-scattered by the synchrotron-producing electrons inside the jet; the jet cyclo-synchrotron emission is also subject to self-Comptonisation (SSC). The jet base serves as the effective corona to re-illuminate the disc and generate the disc reflection spectrum, calculated by model \texttt{reflect} \citep{magdziarz1995angle}. Jet model parameters are listed in Table~\ref{tb:pars}.}\label{fig:model}
\end{figure}

\texttt{Bljet}, the spectral model used in this work to fit the time-independent, broadband spectra (radio through X-rays or $\gamma$-rays) of accreting black holes is an update of the model presented first in \cite{lucchini2019breaking}, and further developed in \cite{lucchini2021}. Briefly, \texttt{bljet} is a steady-state, multi-zone, semi-analytical model, designed to qualitatively mimic the dynamical jet profiles observed in GRMHD simulations, while being computationally inexpensive enough to allow complex multi-wavelength data analysis. The previous version (called \texttt{agnjet}) has mainly been used in studies of hard states and the quiescence of BHXBs \citep[e.g.][]{markoff2005going,gallo2007spectral,connors2017mass}. It can also reproduce the broadband SEDs of several low luminosity AGN \citep[e.g.][]{markoff2001nature,markoff2008results,maitra2011jet,markoff2015above,connors2017mass}. The details of the jet dynamics can be found in \cite{lucchini2019breaking}, while the treatment of the particle distributions and radiative mechanisms are described in \cite{lucchini2021}. 

Fig.~\ref{fig:model} gives a schematic of the model geometry used in this work. Besides a truncated Shakura-Sunyaev accretion disc \citep{shakura1973black,shakura1976theory}, the model parametrises the power injected at the base of the jet as a percentage $N_j$ of the black hole's Eddington luminosity $L_{\rm Edd}$, divided among hot electrons, cold protons and magnetic fields. We only consider leptonic radiative processes (but see also \citealt{kantzas2021new} for a version of the model which includes the contribution of hadronic processes); the protons carried in the jet are always assumed to be cold and therefore do not contribute to the SED. The base of the jets is a cylinder of radius $R_0$ and height $H=2R_0$ located above the accretion disc and at an initial height of $Z_0 = 2\,\rm{R_g}$ above the disc and black hole (although note that the model currently does not include general relativistic effects, and therefore we never fit for $Z_0$). For typical parameters, the emission from the base of the jets resembles that of a typical black hole corona, and therefore our model can be thought of as a physical realisation of the lamp-post model. Hereafter the terms "corona" and "jet base" are used interchangeably. We include the thermal cyclo-synchrotron emission from near the base of the jet, the non-thermal synchrotron emission downstream in the jet, and the inverse-Compton scattering of both disc thermal photons and cyclo-synchrotron photons. The coronal emission then re-illuminates the accretion disc and generates the disc reflection spectrum \citep[e.g.][]{markoff2004constraining,fabian2010x}; in this work, we treat reflection using the phenomenological model \texttt{reflect} \citep{magdziarz1995angle}. Additionally, \texttt{bljet} can also include an optical black body excess to mimic the optical/UV emission from the disc irradiation \citep[e.g.][]{gierlinski2009reprocessing}.

\begin{figure*}
\centering
\subfloat[\label{fig:j1752hid}]{\includegraphics[width=0.48\textwidth]{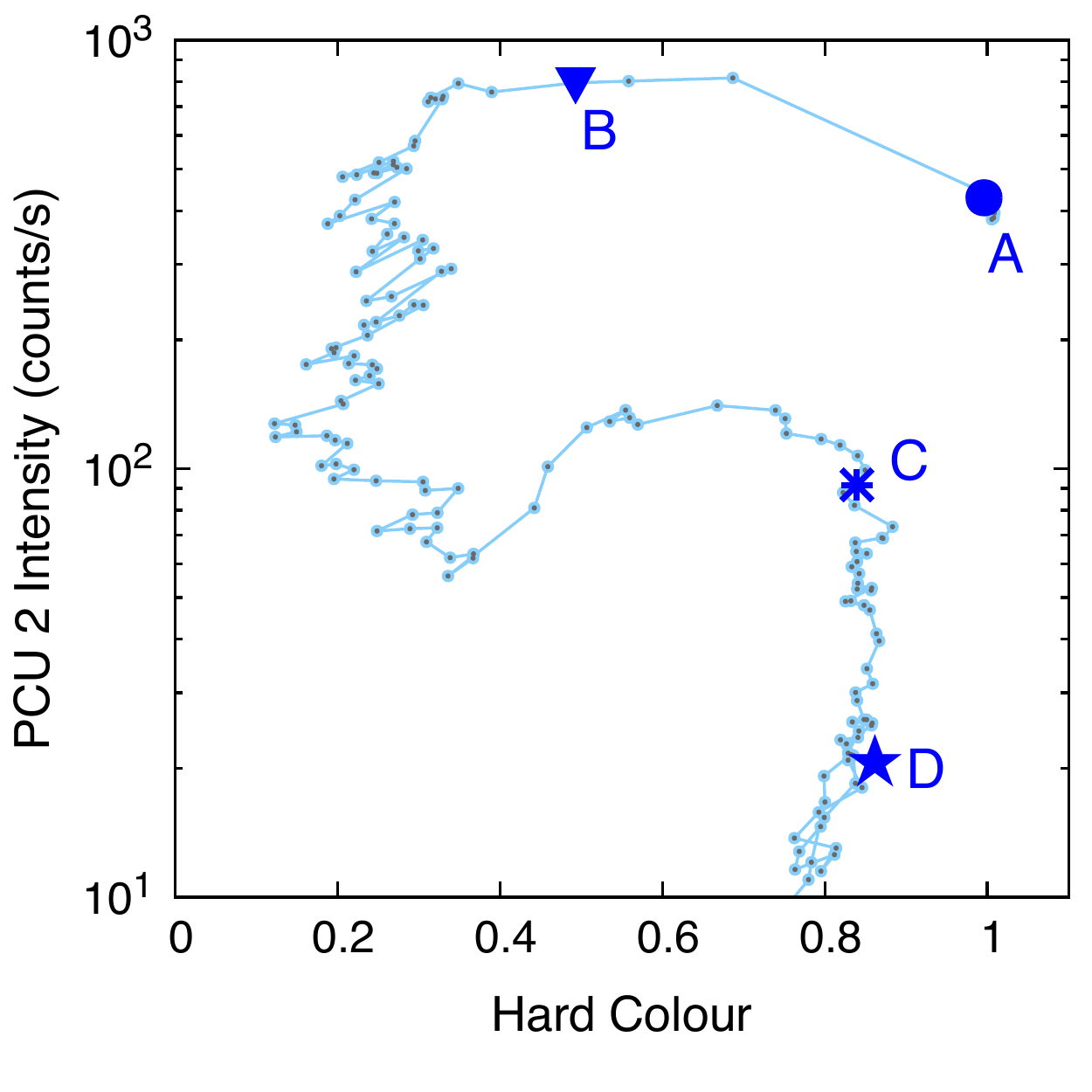}}\hfill 
\subfloat[\label{fig:j1659hid}]{\includegraphics[width=0.48\textwidth]{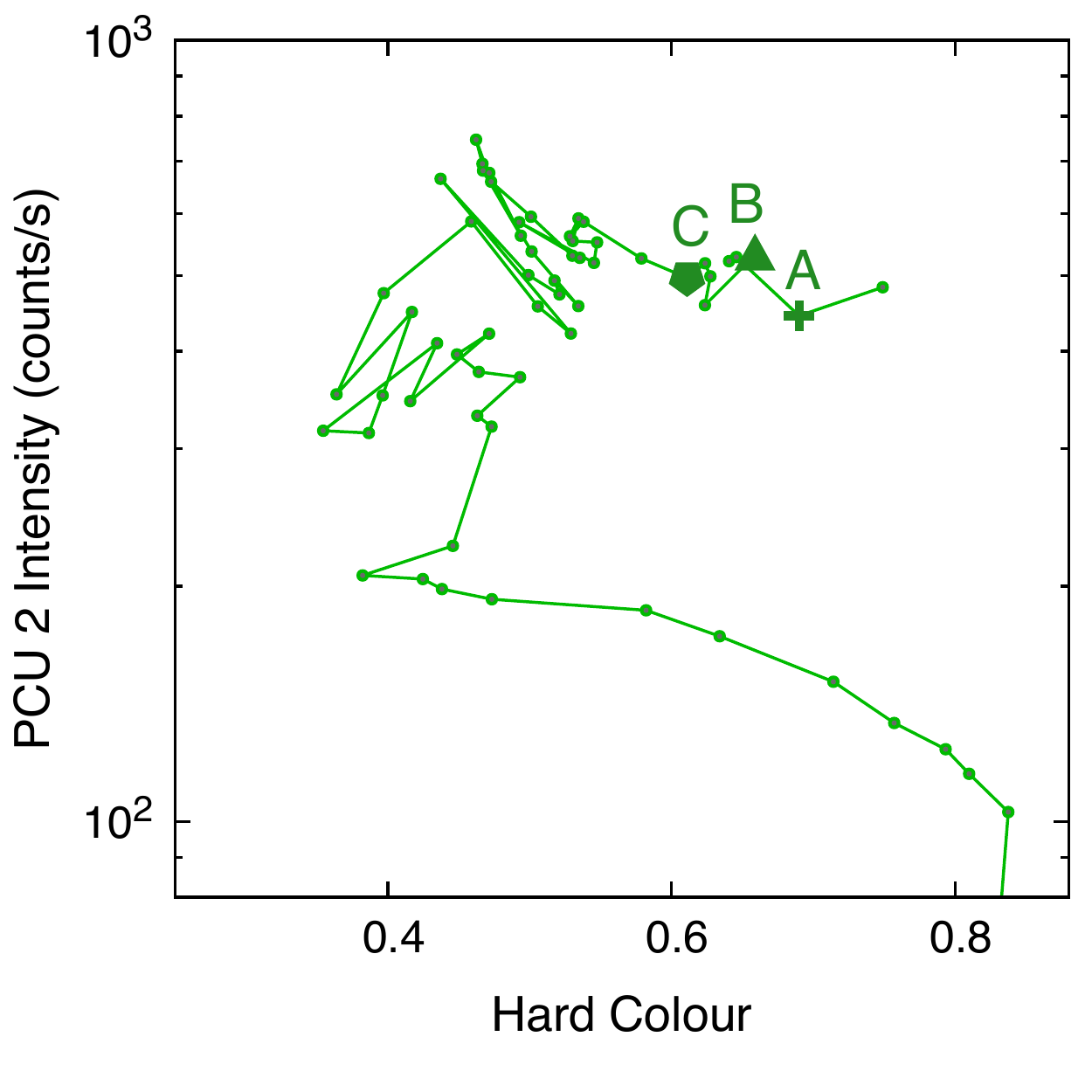}}\hfill
\subfloat[\label{fig:j1650hid}]{\includegraphics[width=0.48\textwidth]{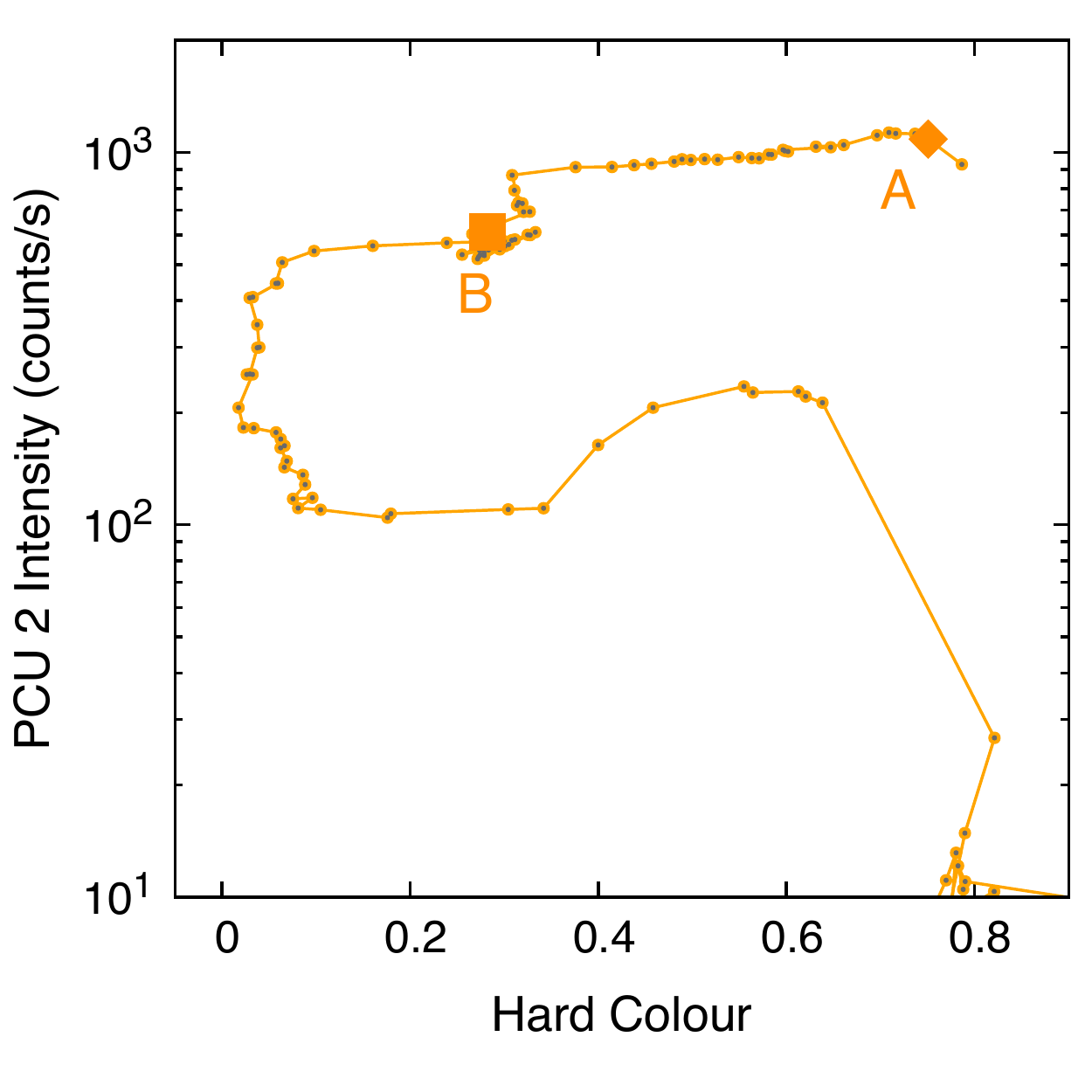}}\hfill
\subfloat[\label{fig:pc3}]{\includegraphics[width=0.48\textwidth]{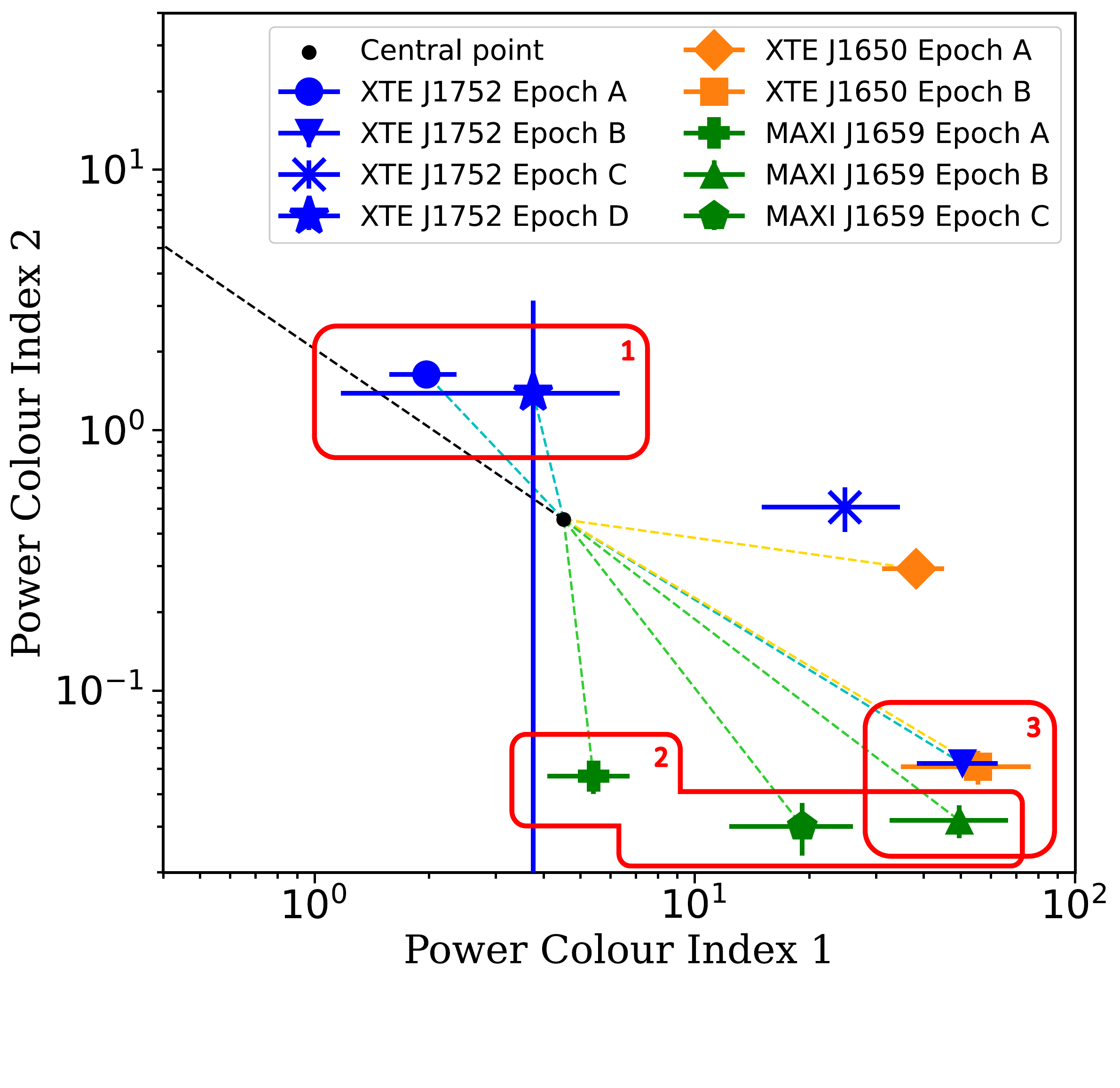}}
\caption{X-ray hardness-intensity diagrams (HIDs) of \textbf{a)} XTE J1752-223, \textbf{b)} MAXI J1659-152, and \textbf{c)} XTE J1650-500, derived from \textit{RXTE}/PCA data. The X-ray hard colour in HIDs is calculated as the ratio of source counts between $8.6-18.0\,\rm{keV}$ and $5.0-8.6\,\rm{keV}$. Quasi-simultaneous multi-wavelength observation epochs highlighted by A through D in each source are listed in the Appendix. \textbf{d)} The power colour diagram (PCD) including all the corresponding epochs (Power Colour Index 1 = variance in 0.25-2.0 Hz/0.0039-0.031 Hz, and Power Colour Index 2 = variance in 0.031-0.25 Hz/2.0-16.0 Hz). Symbols of epochs are consistent with each other in HIDs and PCD. The dashed black line indicates where hue=0$^{\circ}$. The epoch clusters we identified for multi-wavelength jointly-fitting are highlighted by red boxes (see \ref{sec:mwl}). } 
\label{fig:3src}
\end{figure*}

Details of the jet acceleration and collimation profile are discussed in \cite{lucchini2019breaking}. The jet bulk acceleration begins at the top of the corona, effectively converting the initial magnetic field content of the jet into bulk kinetic energy up to a distance $z_{\rm acc}$ from the black hole. At a distance $z_{\rm diss}$, which we take to be equal to $z_{\rm acc}$ for simplicity, the jet experiences a dissipation region, where a fraction $\epsilon_{pl}$=10\% of the thermal electrons are channelled into a non-thermal, power-law particle distribution tail with an index of $p$. This dissipation region represents the start of particle acceleration in the jet \citep{markoff2005going} which sets the start of the flat/inverted spectrum \citep{blandford1979relativistic,boettcher2010timing,malzac2013internal}, possibly due to internal shocks driven by fluctuations in the outflow velocity \citep{malzac2014spectral}, or instabilities along the jet boundary \citep{chatterjee2019accelerating}. As the plasma moves downstream of $z_{\rm acc}$, the percentage of accelerated particles along the jet is reduced by a factor $(\log_{10}(z_{\rm acc})/\log_{10}(z))^{f_{\rm pl}}$. The parameter $f_{\rm pl}$ allows us to artificially suppress the cyclo-synchrotron emissivity from successive jet regions outwards along the jets, resulting in an inverted, rather than flat, radio spectrum. This parameter in effect fudges the complex balance between particle cooling and re-acceleration that we plan to explore in future work. A list of the jet parameters and descriptions concerning this work are listed in Table~\ref{tb:pars}.

We freeze several parameters in our SED modelling in order to reduce model degeneracy. BHXB jets are likely only moderately relativistic \citep{fender2004towards}, and hence we take the terminal bulk Lorenz factor of the jet to be $\Gamma=3$. The final magnetisation parameter $\sigma_{\rm f}$ after the bulk acceleration at $z_{\rm acc}$ is set to be 0.1, as in \cite{lucchini2021}. In our data, the optically thin part of the non-thermal synchrotron spectrum is not well sampled, and therefore we freeze $p$ and $z_{\rm acc}$ to $2.3$ and $2000\,\rm{R_g}$, respectively, in agreement with the estimates by \cite{gandhi2008rapid,gandhi2011} for the canonical BHXB GX 339-4. This choice of $p$ and $z_{\rm acc}$ causes the X-ray flux of the non-thermal synchrotron to fall well below the inverse-Compton emission from the jet base. The standard plasma parameter $\beta_p=U_e/U_B$ at the base of the jet is set to 0.0315. This choice of the $\beta_p$, together with the final Lorenz factor $\gamma_{\rm acc}$ and the final magnetisation $\sigma_{\rm f}$, sets the number ratio between electron-positron pairs and protons to be $\approx 30$, resulting in jet powers that are roughly on the order of the accretion disc luminosity, similarly to \cite{lucchini2021}. A pair-dominated region near the black hole would be a natural consequence of several mechanisms for pair-loading in the black hole a magnetosphere \citep[e.g.][]{neronov2007production,moscibrodzka2011pair,broderick2015horizon}. Other parameter values fixed among fits can be found in Table~\ref{tb:pars} as well.

The two primary geometrical parameters we focus on in this work are the size of the jet base $R_0$, and the radius of the inner edge of the Shakura-Sunyaev disc $R_{\rm in}$. Crucially, $R_0$ scales the energy densities and optical depth of the corona, and thus has a large impact on the flux of the thermal synchrotron emission inside the corona, as well as on the normalisation and slope of the inverse-Compton spectrum.

%%%% RESULTS %%%%%%%%%%%%%%%%%%%%
\section{Results}
\label{sec:results}

\subsection{Overall spectral and timing behaviour}

\begin{table*}
\renewcommand{\arraystretch}{1.5}
\centering
\caption[111]{Best-fitting parameters of the selected epochs with phenomenological model \texttt{TBabs*(reflect(powerlaw)+diskbb+gaussian)}. Parameters fixed to certain values are shown in brackets. Unconstrained parameters are given with their upper limits in 90\% confidence level. \textit{Swift}/XRT and \textit{RXTE}/HEXTE data are included when available. Galactic absorption for each source is fixed to reported values in the literature (XTE J1650-500:\citealt{miniutti2004relativistic}; MAXI J1659-152:\citealt{van2013broad}; XTE J1752-223:\citealt{garcia2018reflection}). Epochs in the same cluster are modelled individually, and J1659B is modelled only once while it is displayed in both Cluster 2 and Cluster 3 for comparison purpose.}
\resizebox{\textwidth}{!}{%
\begin{tabular}{cc|cc|ccc|ccc}
\hline
&&\multicolumn{2}{c|}{Cluster 1}&\multicolumn{3}{c|}{Cluster 2}&\multicolumn{3}{c}{Cluster 3}\\
Model Component&Parameter&J1752 A&J1752 D&J1659 A&J1659 B&J1659 C&J1752 B&J1659 B&J1650 B\\
\hline
TBabs&$N_H$ ($10^{22}cm^{-2}$)&\multicolumn{2}{c|}{(1.0)}&\multicolumn{3}{c|}{(0.32)}&(1.0)&(0.32)&(0.5)\\
\hline
reflect&$\rm{Refl}_{\rm frac}$&$0.40^{+0.03}_{-0.03}$&$<0.79$&$<1\times10^{-5}$&$<0.18$&$<0.05$&$0.19^{+0.19}_{-0.17}$&$<0.18$&$0.7^{+0.4}_{-0.4}$\\
\hline
powerlaw&$\Gamma$&$1.525^{+0.007}_{-0.007}$&$1.735^{+0.003}_{-0.002}$&$2.056^{+0.008}_{-0.008}$&$2.17^{+0.03}_{-0.02}$&$2.137^{+0.025}_{-0.005}$&$2.17^{+0.07}_{-0.07}$&$2.17^{+0.03}_{-0.02}$&$2.4^{+0.2}_{-0.1}$\\
%%no-longer-used&$E_{cut}$ (keV)&$356^{+42}_{-34}$&$(500)$&$(500)$&$(500)$&$(500)$&$(500)$&$(500)$&$>290$\\
&$A_{powerlaw}$&$0.610^{+0.011}_{-0.007}$&$0.048^{+0.004}_{-0.002}$&$1.74^{+0.03}_{-0.03}$&$2.41^{+0.12}_{-0.08}$&$2.00^{+0.01}_{-0.12}$&$2.7^{+0.5}_{-0.4}$&$2.41^{+0.12}_{-0.08}$&$1.1^{+0.4}_{-0.3}$\\
\hline
diskbb&$T_{in}$ (keV)&$0.28^{+0.03}_{-0.03}$&$0.16^{+0.04}_{-0.04}$&$0.52^{+0.02}_{-0.02}$&$0.61^{+0.02}_{-0.02}$&$0.772^{+0.016}_{-0.005}$&$0.76^{+0.02}_{-0.03}$&$0.61^{+0.02}_{-0.02}$&$0.661^{+0.008}_{-0.008}$\\
&$A_{diskbb}$&$1.4^{+1.3}_{-0.6}\times10^{4}$&$5^{+79}_{-4}\times10^{4}$&$1.1^{+0.2}_{-0.2}\times10^{3}$&$7.6^{+1.1}_{-0.9}\times10^{2}$&$4.0^{+0.5}_{-0.1}\times10^{2}$&$2.0^{+0.3}_{-0.3}\times10^{3}$&$7.6^{+1.1}_{-0.9}\times10^{2}$&$7.1^{+0.5}_{-0.5}\times10^{3}$\\
\hline
gaussian&$E_{center}$ (keV)&\multicolumn{2}{c|}{(6.4)}&\multicolumn{3}{c|}{(6.4)}&\multicolumn{3}{c}{(6.4)}\\
&$\sigma_{gauss}$&$<0.6$&$<1.1$&$<0.01$&$<0.6$&$1.1^{+0.4}_{-0.3}$&$0.91^{+0.08}_{-0.07}$&$<0.6$&$1.10^{+0.02}_{-0.02}$\\
&$A_{gauss}$&$5^{+4}_{-4}\times10^{-4}$&$3^{+2}_{-2}\times10^{-4}$&$9^{+4}_{-4}\times10^{-4}$&$10^{+5}_{-5}\times10^{-4}$&$0.004^{+0.001}_{-0.001}$&$0.020^{+0.003}_{-0.003}$&$10^{+5}_{-5}\times10^{-4}$&$0.015^{+0.001}_{-0.001}$\\
\hline 
$\chi^2$/d.o.f&&277/174&35/53&195/120&304/179&159/100&79/38&304/179&74/51\\
\hline
\end{tabular}}
\label{tb:bbporefl}
\end{table*}

Fig.~\ref{fig:3src} shows the hardness-intensity diagrams (HIDs) and the power-colour diagram (PCD) of outbursts from our selected BHXBs: XTE J1752-223, MAXI J1659-152, and XTE J1650-500. The X-ray hard colour in the HIDs is calculated as the ratio of source counts between $8.6-18.0\,\rm{keV}$ and $5.0-8.6\,\rm{keV}$. All epochs with quasi-simultaneous observations are labelled with letters in the HIDs and are shown in the PCD. To illustrate the chronology of the selected multi-wavelength epochs, we use the HID to map the full outburst of each source, and the data reduction to produce the HIDs is identical to that used in \cite{connors2020evidence}. 

\begin{figure}
    \centering
    \includegraphics[width=\linewidth]{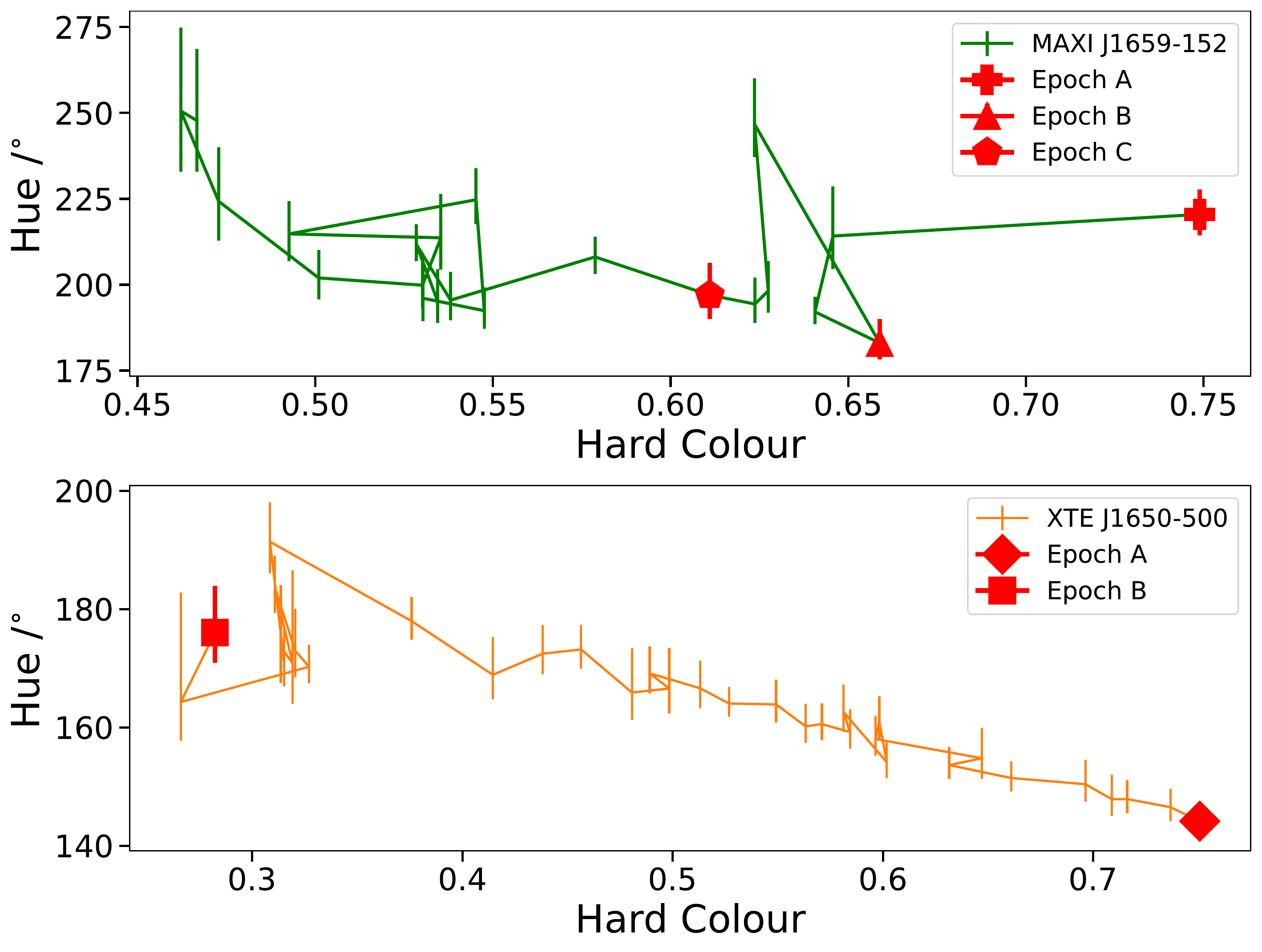}
\caption{Evolution of the power spectral hue in MAXI J1659-152 and XTE J1650-500 with the hardness, focused on observations close to the selected epochs in this work. Epochs A through C are highlighted by red data points for each source. During the concerned period of outburst, the hue of MAXI J1659-152 changed rapidly, and does not appear to strongly follow the spectral changes; XTE J1650-500 showed a more prominent trend of the hue increasing as the source transited from hard states to soft states over time.}\label{fig:j1659j1650hue}
\end{figure}

\begin{figure}
    \centering
    \includegraphics[width=\linewidth]{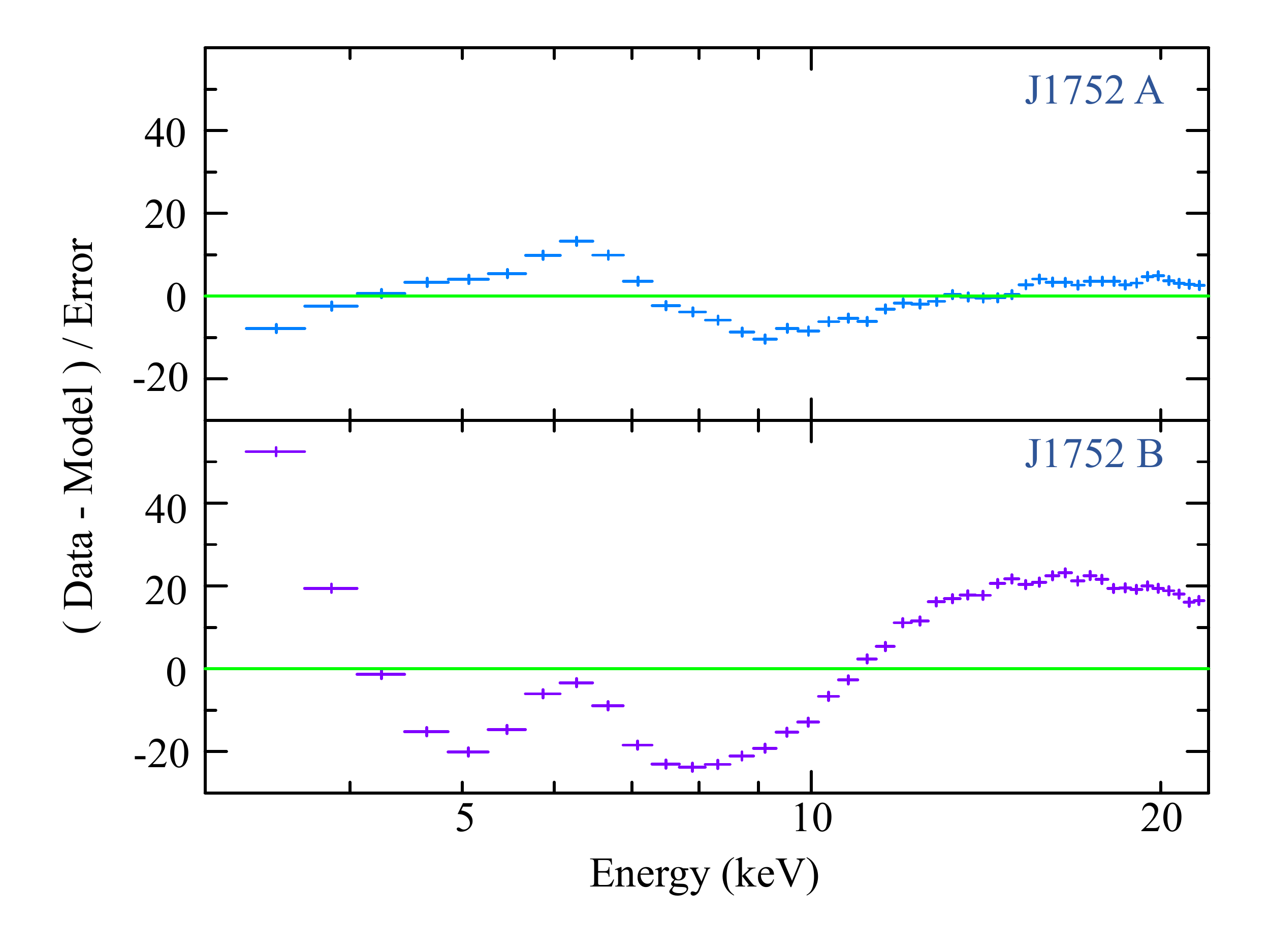};
\caption{Residuals of XTE J1752-223 Epoch A (2009-11-05) and B (2010-01-21) \textit{RXTE}/PCA data fitted with a power-law model. The slope and normalisation of the model are different between these two epochs. Both epochs show the broad iron K-$\alpha$ emission at 6.4 $\rm keV$ and the Compton hump $\gtrsim10$ $\rm keV$, indicating the presence of the disc reflection. Moreover, J1752B shows a powerful soft component that comes from the thermal disc emission, signaling the ongoing state transition (Fig.~\ref{fig:j1752hid}). }\label{fig:j1752po}
\end{figure}

We identify three epoch clusters of interests from the power-colour diagram, and focus our spectral analyses of these clusters by performing joint spectral fits of the broadband SEDs for each (Fig.~\ref{fig:pc3}): Cluster 1 consists of XTE J1752-223 Epochs A\&D, which have similar hues; these are from a bright hard state during the rise of the outburst, and a faint hard state during the decay, respectively. Cluster 2 consists of MAXI J1659-152's Epochs A\&B\&C, capturing the gradual changes in spectral and timing properties during the hard-to-soft state transition of the source. Cluster 3 consists of bright hard-intermediate states from all three objects (XTE J1752-223 Epoch B, MAXI J1659-152 Epoch B, and XTE J1650-500 Epoch B), allowing us to probe states showing near identical hues but different spectral characteristics. Although XTE J1752-223 on 2010-01-21 (Epoch B) only has X-ray and radio observations, we find its power-colour hue is consistent with two epochs from the other two sources (J1650B and J1659B), and so we make an exception of the selection criteria and keep this epoch in our following multi-wavelength spectral analysis in order to exploit the X-ray timing similarity across sources. On the other hand, XTE J1752-223 Epoch C and XTE J1650-500 Epoch A have hue measurements that differ from each other at the $\approx 2-3\sigma$ level, and furthermore, they happen at different outburst stages from different sources. This makes it harder to directly compare the two, and thus we do not form an additional cluster containing these two epochs. We only proceed with the clustered epochs in the following spectral analyses.

Our findings of the power colours in three BHXB sources are mostly consistent with the previous study on 12 BHXB systems \citep{heil2015power}: epochs with states of different spectral hardness occupy different angular regions on the power colour diagram. We notice that the three selected MAXI J1659 epochs show a type-C Quasi-Periodic Oscillation (QPO) in their power spectra \citep{kalamkar2011identification}, which combined with high inclination (>$60^{\circ}$, also see the Appendix for a discussion on the inclination estimation of J1659) can alter the evolution track of the BHXB source on the PCD (Fig.~\ref{fig:pc}) and mainly push the track to have lower PC2 values in the concerned range of hue for these three epochs \citep{heil2015inclination}. However, for the purposes of our study this effect is minimal and does not affect our selection of epoch clusters. Therefore, in this work we do not remove any QPOs before calculating the hue. In addition, we find that the hue can also fluctuate in individual PCD tracks while the sources become softer in X-rays (Fig.~\ref{fig:j1659j1650hue}), blurring the generally increasing trend of hue (moving clockwise on the PCD) at these outburst stages. These results show that some additional caution when interpreting the empirical spectral-timing correlation in individual BHXB sources should be taken.

Our next step is to perform spectral analyses of the epochs in each of the three clusters. Before applying the multi-wavelength jet model, we perform phenomenological modelling to the X-ray data in order to quantify the amount of reflection in each epoch, as this is a very important constraint on jet models: as a rule of thumb, low or absent reflection fraction and/or narrow iron K--$\alpha$ lines favour non-thermal synchrotron emission as the origin of the X-ray power-law, while large reflection fractions and/or broad lines favour inverse-Compton emission near the jet base \citep{markoff2004constraining}. The fitting and statistical analysis of the phenomenological modelling is carried out using the \texttt{XSPEC} package (version 12.11.1; \citealt{arnaud1996xspec}). By using a simple power-law model, we identify broad iron K-$\alpha$ emission and the Compton hump in the fit residuals of several epochs, indicating the presence of relativistic disc reflection. Examples of the fitting residuals can be found in Fig.~\ref{fig:j1752po}. We then use the phenomenological model \texttt{TBabs*(reflect(powerlaw)+diskbb+gaussian)} to model the X-ray data of all three clusters. We use the model \texttt{reflect} \citep{magdziarz1995angle} plus a Gaussian line profile fixed at 6.4 keV to model the disc reflection features, while \texttt{diskbb} covers the disc thermal photons. We set the iron abundance of the disc to the solar abundance, except for XTE J1752, in which it is set to 3.5 times the solar abundance \citep{garcia2018reflection}. Best-fitting parameters are summarised in Table~\ref{tb:bbporefl}. We note that for the two epochs in Cluster 1 the disc normalisation is abnormally high compared to other epochs. This difference is because the disc temperature $T_{in}$ is lower in these two epochs, making the disc component constrained only by the soft end of the \textit{Swift}/XRT data. Combined with our choice to fix the absorption $N_H$, this makes our phenomenological fits only a rough estimation on the disc component and should be interpreted with caution. We also observe the disc component falling almost out of the X-ray band in our multi-wavelength SED modelling of these two epochs (Fig.~\ref{fig:mwl1}).

\subsection{Multi-wavelength SEDs}
\label{sec:mwl}

\begin{figure}
\centering
\includegraphics[width=\linewidth]{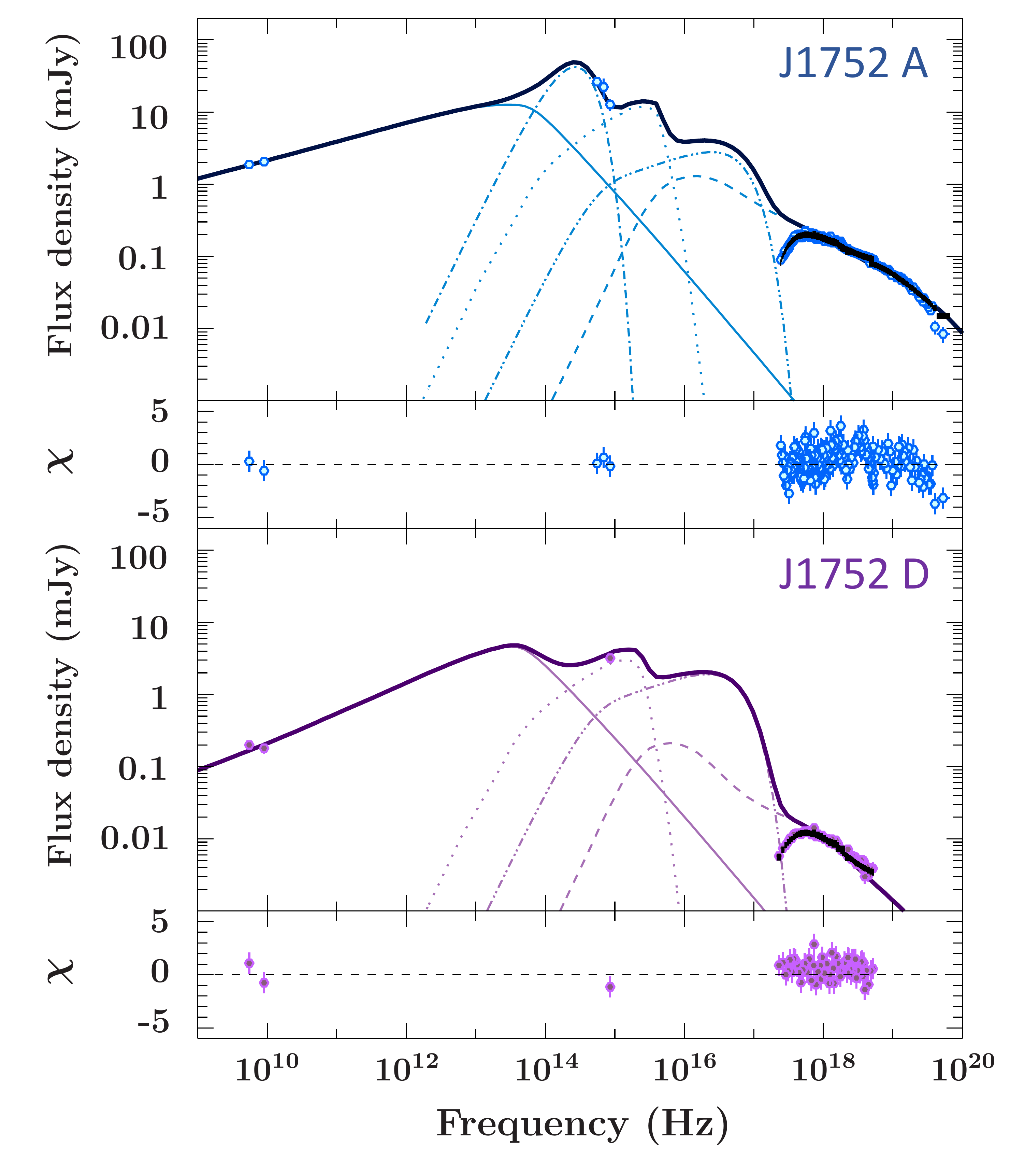}
\caption{Best joint-fit of the two epochs in Cluster 1. The thick continuous dark line shows the total model, the thin continuous line shows the non-thermal synchrotron emission, the thin dashed line the inverse-Compton emission from the jet base, the thin dotted line shows the thermal cyclo-synchrotron emission from the jet base, the dot-dashed and double dot-dashed lines show the optical black body excess and accretion disc, the triple dot-dashed line represents reflection.} \label{fig:mwl1}
\end{figure}

\begin{figure}
\centering
\includegraphics[width=\linewidth]{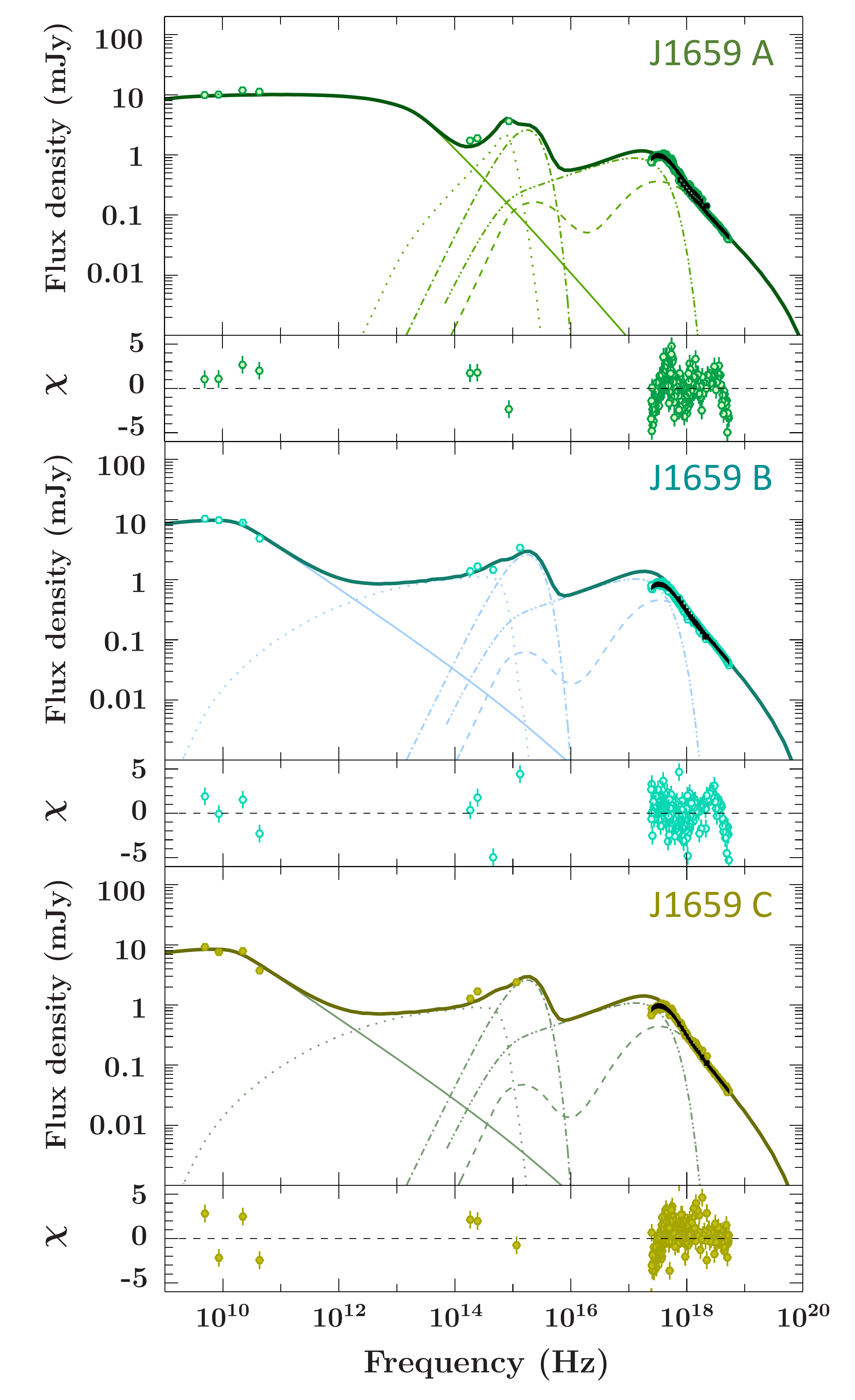}
\caption{Best joint-fit of SEDs in Cluster 2. Same format as Fig.~\ref{fig:mwl1}.} \label{fig:mwl2}
\end{figure}

\begin{figure}
\centering
\includegraphics[width=\linewidth]{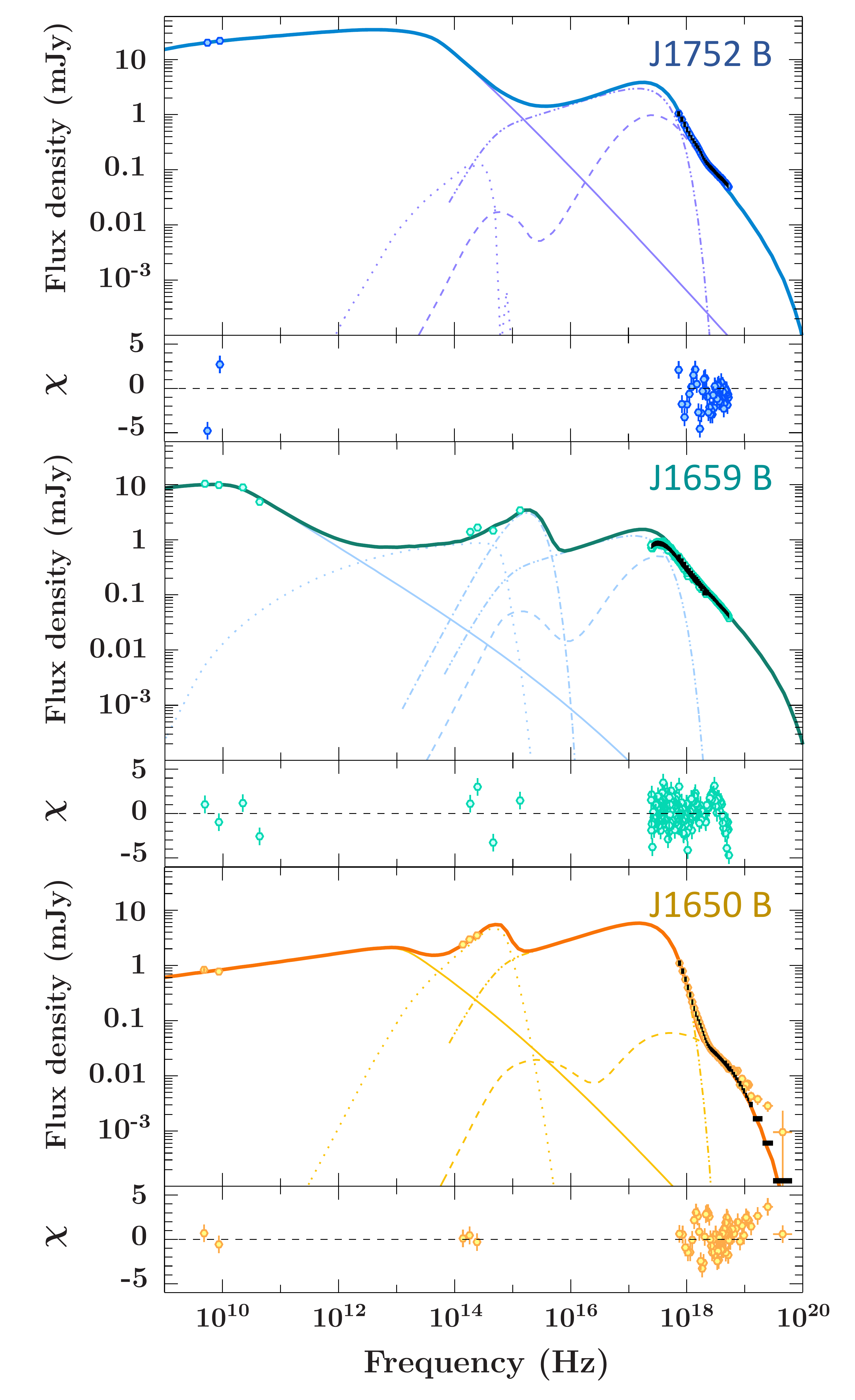}
\caption{Best joint-fit of SEDs in Cluster 3. Same format as Fig.~\ref{fig:mwl1}, with y-axis range changed to cover the X-ray high-energy tail of J1650B.} \label{fig:mwl3}
\end{figure}

\begin{table*}
\centering
\caption[List of known parameters for selected sources]{Fixed parameters of the sources in this work. $M_{\rm BH}$ is the black hole mass in units of $M_{\odot}$, $d$ and $\theta_i$ the distance and inclination angle of the system, and $N_H$ the hydrogen column density for interstellar extinction. We freeze all of these values in our spectral fits.}
\begin{tabular}{cccccc}
\hline
 & $M_{\rm BH}$ ($M_\odot$) & $d$ ($kpc$) & $\theta_i$ ($^{\circ}$) & $N_H$ ($10^{22}cm^{-2}$) & References\\
\hline
XTE J1752-223 & 9.6 & 6 & 35 & 1.0 & 1,2,3\\
MAXI J1659-152 & 6 & 6 & 75 & 0.32 & 4,5,6,7\\
XTE J1650-500 & 5.1 & 2.6 & 45 & 0.5 & 8,9,10\\
\hline
\end{tabular}
\label{tb:srcpars}
\begin{tablenotes}
\item[1]1: \cite{shaposhnikov2010discovery}; 2: \cite{garcia2018reflection}; 3: \cite{ratti2012black}; 4: \cite{molla2016estimation}; 5: \cite{kuulkers2013maxi}; 6: \cite{kong2012optical}; 7: \cite{van2013broad}; 8: \cite{slany2008mass}; 9: \cite{miniutti2004relativistic}; 10: \cite{homan2006xmm}.
\end{tablenotes}
\end{table*}

\begin{table*}
\renewcommand{\arraystretch}{1.5}
\centering
\caption[111]{Best-fitting parameters of the joint-fits in this work. Fixed parameters are shown in brackets. Unconstrained parameters are given with their upper limits in 90\% confidence level. Parameters tied to each other within each cluster of epochs are indicated via a shared column. All epochs use a fixed value of 2000 $R_g$ as $z_{acc}$ except MAXI J1659 Epoch B\&C have a fixed $z_{acc}$ of $4\times10^{6}$ $R_g$ in order to model the radio spectral break observed in these two epochs. Because of the lack of UV data in each of the J1659 epochs, we combine three epochs in Cluster 2 and model the UV excess phenomenologically using one black body component with a fixed temperature (30000K). Both Cluster 2 and Cluster 3 have epoch J1659B thus we include this epoch in the joint-fits respectively.}
\resizebox{\textwidth}{!}{%
\begin{tabular}{cc|cc|ccc|ccc}
\hline
&&\multicolumn{2}{c|}{Cluster 1}&\multicolumn{3}{c|}{Cluster 2}&\multicolumn{3}{c}{Cluster 3}\\
Model Component&Parameter&J1752 A&J1752 D&J1659 A&J1659 B&J1659 C&J1752 B&J1659 B&J1650 B\\
\hline
TBabs&$N_H$ ($10^{22}cm^{-2}$)&\multicolumn{2}{c|}{(1.0)}&\multicolumn{3}{c|}{(0.32)}&(1.0)&(0.32)&(0.5)\\
\hline
Bljet%&$M_{BH}$ ($M_{\odot}$)&\multicolumn{2}{c|}{(9.6)}&\multicolumn{3}{c|}{(6)}&(9.6)&(6)&(5.1)\\
%&$Incl$ ($^{\circ}$)&\multicolumn{2}{c|}{(35)}&\multicolumn{3}{c|}{(75)}&(35)&(75)&(45)\\
%&$d$ ($kpc$)&\multicolumn{2}{c|}{(6)}&\multicolumn{3}{c|}{(6)}&(6)&(6)&(2.6)\\
&$f_{\rm pl}$&$8.6^{+0.2}_{-0.4}$&$15^{+1}_{-1}$&\multicolumn{3}{c|}{(0)}&$2.94^{+0.04}_{-0.17}$&(0)&$3.3^{+0.3}_{-0.2}$\\
&$N_j$&$0.024^{+0.001}_{-0.002}$&$0.014^{+0.001}_{-0.002}$&$0.0395^{+0.0008}_{-0.0006}$&$0.0324^{+0.0007}_{-0.0006}$&$0.0294^{+0.0006}_{-0.0004}$&$0.0374^{+0.0005}_{-0.0015}$&$0.0335^{+0.0006}_{-0.0006}$&$0.00316^{+0.00007}_{-0.00007}$\\
&$R_0$ ($R_g$)&\multicolumn{2}{c|}{$12.3^{+0.5}_{-0.9}$}&$40.5^{+0.7}_{-0.7}$&$33.0^{+0.4}_{-0.4}$&$32.6^{+0.6}_{-0.6}$&\multicolumn{3}{c}{$34.6^{+0.4}_{-0.5}$}\\
&$T_e$ ($keV$)&$242^{+12}_{-7}$&$188^{+18}_{-17}$&$159^{+3}_{-4}$&$110^{+3}_{-4}$&$107^{+3}_{-4}$&$44^{+2}_{-2}$&$100^{+3}_{-4}$&$324^{+5}_{-5}$\\
&$R_{\rm in}$ ($R_g$)&\multicolumn{2}{c|}{$140^{+13}_{-16}$}&\multicolumn{3}{c|}{$20.0^{+0.2}_{-0.2}$}&$9.9^{+0.6}_{-0.3}$&$24.0^{+0.4}_{-0.4}$&$12.3^{+0.2}_{-0.1}$\\
&$L_{disc}$ ($L_{Edd}$)&$0.011^{+0.002}_{-0.003}$&$0.007^{+0.001}_{-0.001}$&$0.087^{+0.001}_{-0.001}$&$0.107^{+0.001}_{-0.002}$&$0.114^{+0.001}_{-0.002}$&$0.069^{+0.002}_{-0.001}$&$0.113^{+0.002}_{-0.002}$&$0.0546^{+0.0006}_{-0.0006}$\\
\hline
reflect&$\rm{Refl}_{\rm frac}$&$0.29^{+0.02}_{-0.02}$&$0.5^{+0.2}_{-0.2}$&$<5\times10^{-4}$&$<4\times10^{-4}$&$<3\times10^{-4}$&$0.41^{+0.06}_{-0.05}$&$0.03^{+0.03}_{-0.02}$&$<8\times10^{-4}$\\
\hline
gaussian&$E_{center}$ (keV)&\multicolumn{2}{c|}{(6.4)}&\multicolumn{3}{c|}{(6.4)}&\multicolumn{3}{c}{(6.4)}\\
&$\sigma_{gauss}$&$0.2^{+0.2}_{-0.2}$&$0.004^{+0.014}_{-0.003}$&$0.7^{+0.2}_{-0.1}$&$0.6^{+0.2}_{-0.1}$&$0.4^{+0.4}_{-0.3}$&$0.85^{+0.04}_{-0.04}$&$0.15^{+0.19}_{-0.10}$&$1.10^{+0.02}_{-0.02}$\\
&$A_{gauss}$&$9^{+2}_{-3}\times10^{-4}$&$1.7^{+0.9}_{-0.8}\times10^{-4}$&$3.1^{+0.5}_{-0.5}\times10^{-3}$&$3.1^{+0.5}_{-0.5}\times10^{-3}$&$1.4^{+0.6}_{-0.4}\times10^{-3}$&$18.4^{+0.8}_{-0.8}\times10^{-3}$&$1.3^{+0.3}_{-0.2}\times10^{-3}$&$2.00^{+0.06}_{-0.05}\times10^{-2}$\\
\hline
Black Body&$T_{bbody}$ ($K$)&$4.6^{+1.0}_{-0.8}\times10^{3}$&-&\multicolumn{3}{c|}{(30000)}&-&(30000)&-\\
&$L_{bbody}$ ($erg/s$)&$2.5^{+1.5}_{-0.6}\times10^{36}$&-&\multicolumn{3}{c|}{$1.01^{+0.05}_{-0.04}\times10^{36}$}&-&$1.19^{+0.07}_{-0.06}\times10^{36}$&-\\
\hline 
$\chi^2$/d.o.f&&\multicolumn{2}{c|}{382/240}&\multicolumn{3}{c|}{1909/671}&\multicolumn{3}{c}{766/337}\\
\hline
\end{tabular}}
\label{tb:mwl}
\end{table*}

We then model the broadband SEDs using \texttt{bljet} to explore possible physical similarities among epochs and epoch clusters. To better constrain the model, we test these similarities by performing joint-fits for each cluster, where we model the epochs in one cluster simultaneously and search for best-fits with certain parameters tied between epochs. We use the Interactive Spectral Interpretation System (\texttt{ISIS}, version 1.6.2-43; \citealt{houck2000isis}) to perform our broadband spectral fits, because it allows the forward-folding of the full jet model into X-ray detector space while simultaneously fitting the radio, IR, optical and UV data. We explore the parameter space of our model by using the \textit{emcee} Markov Chain Monte Carlo algorithm \citep{foreman2013emcee} implementation in \texttt{ISIS}. We run the \textit{emcee} for a total of 5000 steps and discard the first 2000 steps, which in all of our cases ensures the convergence of the chain. We use 20 walkers per free parameter for each chain and we define the value of the best-fitting parameter as the median of the walker distribution after the chain burn-in period. We define the 1-$\sigma$ uncertainty as the interval in the posterior distribution which contains 68\% of the walkers after excluding the burn-in period. 

We fix the mass \textbf{$M_{\rm BH}$}, inclination angle \textbf{$\theta_i$}, and distance \textbf{$d$} of the black hole to values reported in the literature, summarised in Table~\ref{tb:srcpars}. The syntax of the multi-wavelength model is "\texttt{constant*TBabs*(reflect*bljet+gaussian)}", where the calibration factors between different instruments are included via the \texttt{constant} model within \texttt{ISIS}. Following \cite{lucchini2021}, we try to tie the geometric parameters of the X-ray emitting region ($R_0$ and $R_{\rm in}$) in each of our epoch clusters when running our joint-fits. The best-fitting results are summarised in Fig.~\ref{fig:mwl1}, Fig.~\ref{fig:mwl2}, Fig.~\ref{fig:mwl3}, and Table \ref{tb:mwl}. We could not find any satisfactory fits with both $R_0$ and $R_{\rm in}$ tied in Cluster 2 or Cluster 3, and instead we find good fits with either $R_0$ (Cluster 2) or $R_{\rm in}$ (Cluster 3) tied in these two joint-fits. Because J1659B is shared by Cluster 2 and Cluster 3, we model this epoch twice in the joint-fits of these two cluster. While the best-fitting parameters of J1659B in these two joint-fits are only marginally consistent with each other, this minor discrepancy does not impact the evolution of $R_0$ or $\rm{Refl}_{\rm frac}$ we infer from the fits (see Section \ref{sec:discussion}). We also check our approach of tying $R_0$ and/or $R_{\rm in}$ parameters by re-doing the joint-fits but with the $R_0$ and $R_{\rm in}$ un-tied among all the epochs. We find that $R_0$ changes less than 1$\sigma$ in most epochs (except for J1659B in Cluster 2 it is less than 3$\sigma$), while $R_{\rm in}$ changes $\sim20\%$ in Cluster 1 and Cluster 2 (see Table \ref{tb:mwl_a} in the Appendix). We discuss various sources of the $R_{\rm in}$ uncertainty in our analyses in the next section.

In all epochs, our model is in good agreement with the broadband SEDs. None of the epochs has sufficient data to constrain the location of the jet spectral break \citep[e.g.][]{corbel2002near,russell2013jet}, while for MAXI J1659-152 the break is only indicated by four radio bands in two epochs (J1659 B\&C), indicating that the particle acceleration region is very far from the black hole \citep{russell2013jet}. The dominant mechanism responsible for the optical to UV band in J1752D and J1650B is the thermal synchrotron emission in the jet base, while four other epochs (J1752A and J1659A/B/C) display an excess in flux (dominant in J1752A) which cannot be fully covered by the coronal synchrotron emission. We model the excess by an additional black body component. One of the possible explanation for this thermal excess is irradiation of the outer disc from the inner X-ray emission \citep[e.g.][]{maitra2009constraining,gierlinski2009reprocessing,van2013broad}. The dominant mechanism for the X-ray emission in our model is inverse-Comptonisation of jet synchrotron photons and disc photons. The up-scattered photons seeded from the disc black body emission dominates the hard X-ray flux except for J1752 Epoch A and Epoch D. Synchrotron-self-Comptonisation (SSC) in the jet dominates the X-ray spectra in the latter two epochs (Fig.~\ref{fig:mwl1}).

%%%% DISCUSSION %%%%%%%%%%%%%%%%%%%%
\section{Discussion}
\label{sec:discussion}

The main results of our joint-fits are as follows: in Cluster 1, our joint-fit models the broadband differences between the high/low HS of J1752 indicating a change in the jet while the accretion disc remains essentially the same. We find that the drop in total luminosity between J1752A and J1752D is caused primarily by three factors: a decrease in the jet power, a decrease in the temperature of the electrons, and an increase in the $f_{\rm pl}$ parameter (implying less efficient particle re-acceleration throughout the jet). Additionally, our results suggest that at the bright hard state the disc is truncated (at $R_{\rm in} = 140^{+13}_{-16} R_g$) as it is in the faint HS with similar hardness ratio (Fig.~\ref{fig:j1752hid}). On the other hand, \cite{garcia2018reflection} finds a $R_{\rm in} = 1.7\pm0.4 R_g$ with detailed reflection spectroscopy using the reflection model \texttt{relxillCp} \citep{dauser2014role} and all of the long-stable high HS data, including the epoch we label J1752A here (but see \citealt{zdziarski2021does} for a truncation radius of $R_{\rm in} \gtrsim 100 R_g$ estimated at the same HS, with which our results are more consistent). An important caveat of the truncation radii estimated in this work is that the lack of UV coverage leads to the poor constraint of the disc thermal continuum. Instead, our estimates rely on matching the normalisation of the Comptonised continuum, which requires computing the energy density of the disc photons as seen by the corona: the inverse-Compton luminosity scales as: $L_{\rm IC}\propto U_{\rm disc} \propto L_{\rm disc}/R^{2}_{\rm in}$, and therefore we can constrain $R_{\rm in}$ only indirectly. Additionally, part of the power-law continuum is due to synchrotron-self Compton, rather than thermal Comptonisation of disc photons. As a result, our numbers for $R_{\rm in}$ should be interpreted with some caution. 

In Cluster 2, during intermediate states towards the HIMS/SIMS transition, we find that the jet in MAXI J1659-152 is decreasing both in power and radius of its base, with the disc parameters staying relatively steady. This is roughly consistent with a contracting corona instead of a gradual reduction of the disc truncation radius, which takes place before the state transition \citep[e.g.][]{fabian2014determination,garcia2015x,kara2019corona}. This coronal behaviour is also suggested by recent X-ray variability studies on other similar sources \citep[e.g.][]{karpouzas2020comptonizing,karpouzas2021variable,garcia2021two}. Noticeably, Cluster 2 contains three epochs with order-of-magnitude PC1 differences (Fig.~\ref{fig:pc3}). However, we don't expect our results to be much affected, because it is the hue rather than a single power colour that tracks the outburst evolution \citep{heil2015power}, and Fig.~\ref{fig:j1659j1650hue} shows that the hue for all three epochs in Cluster 2 are close to each other, within 2$\sigma$ error range. Therefore, it justifies our choice to jointly fit the epochs that they are at physically similar states. Our joint-fit suggests that the disc is moderately truncated during the transition, implying that the disc inner edge should reduce to the innermost circular orbit (ISCO) \citep{novikov1973black} only when the source reaches the SIMS/soft state and the jet is quenched.

In Cluster 3, we explore the spectral similarities among multiple BHXB sources sharing nearly identical hues (Fig.~\ref{fig:pc3}). We find a good fit in which the size of the corona $R_0$ is tied among epochs, and all three epochs are dominated by the inverse-Comptonisation of disc photons. Beyond finding the similarity in the X-ray emitting regions, our fits point to the systems being somewhat different despite sharing similar hue or PSD shapes. In particular, we find that $R_{\rm in}$ needs to be un-tied in order to correctly reproduce the data in this cluster. This is not unexpected given how different the SEDs, and particularly the X-ray spectra, are from each other. This joint-fit provides evidence that the similarity of the PSD/hue among the BHXB systems is driven by a shared coronal geometry, while the disc truncation radius may not be related to the power colours.

\begin{figure}
\centering
\includegraphics[width=\linewidth]{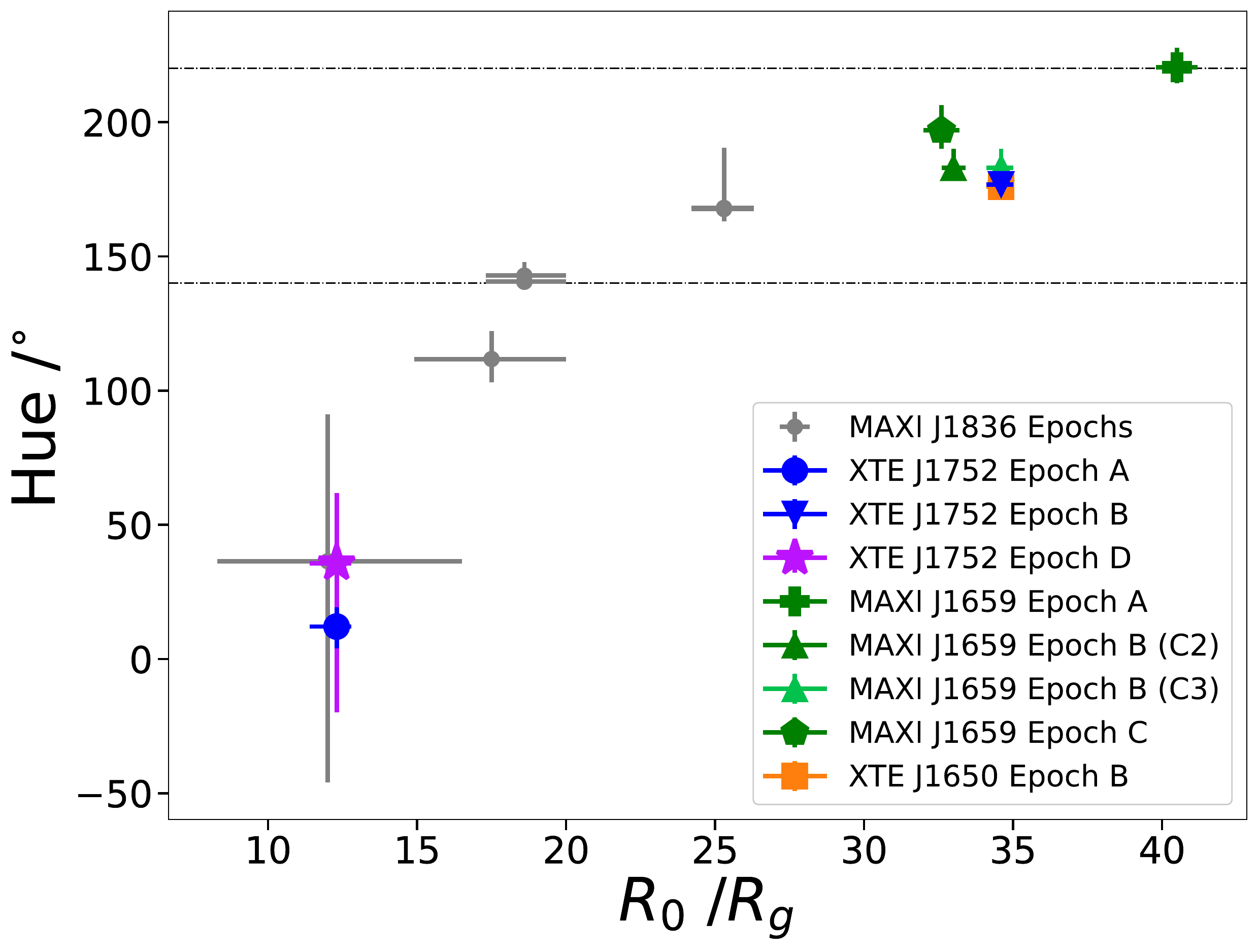}
\caption[$R_0$ vs. hue relation in XTE J1752-223, MAXI J1659-152, XTE J1650-500, and MAXI J1836-194]{$R_0$ vs. hue from the joint-fits of this work (coloured) and from \cite{lucchini2021}, who modelled the outburst of MAXI J1836-194 (grey). Dashed lines indicate the hue ranges corresponding to (from bottom to top): HS, HIMS, and SIMS \citep{heil2015power}. The best-fit $R_0$ of J1659B from both the joint-fits of Cluster 2 (C2, dark green) and Cluster 3 (C3, light green) are plotted.}\label{fig:r0vshue}
\end{figure}

\begin{figure*}
\centering
\subfloat[\label{fig:refl0}]{\includegraphics[width=0.5\textwidth]{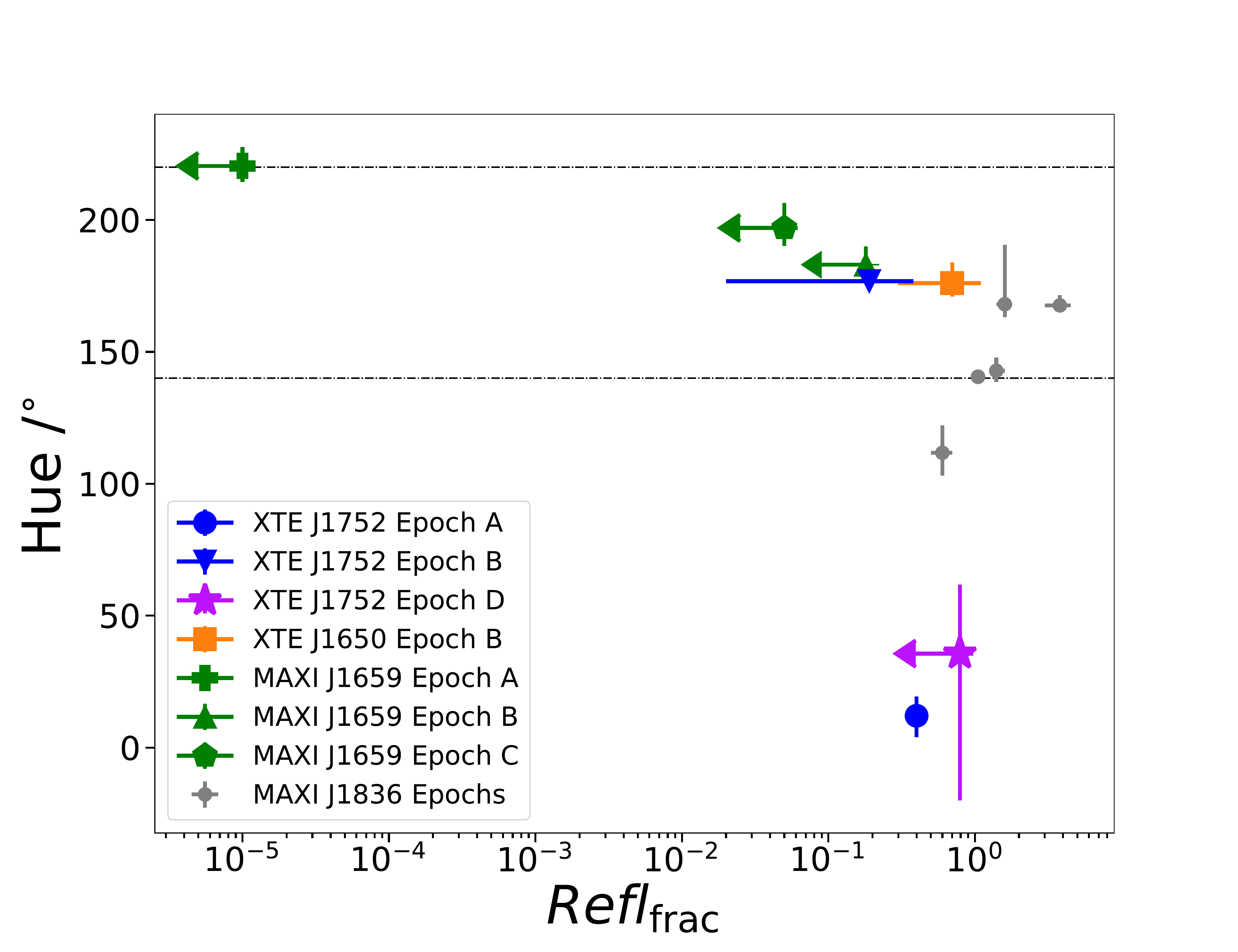}}\hfill
\subfloat[\label{fig:refl1}]{\includegraphics[width=0.5\textwidth]{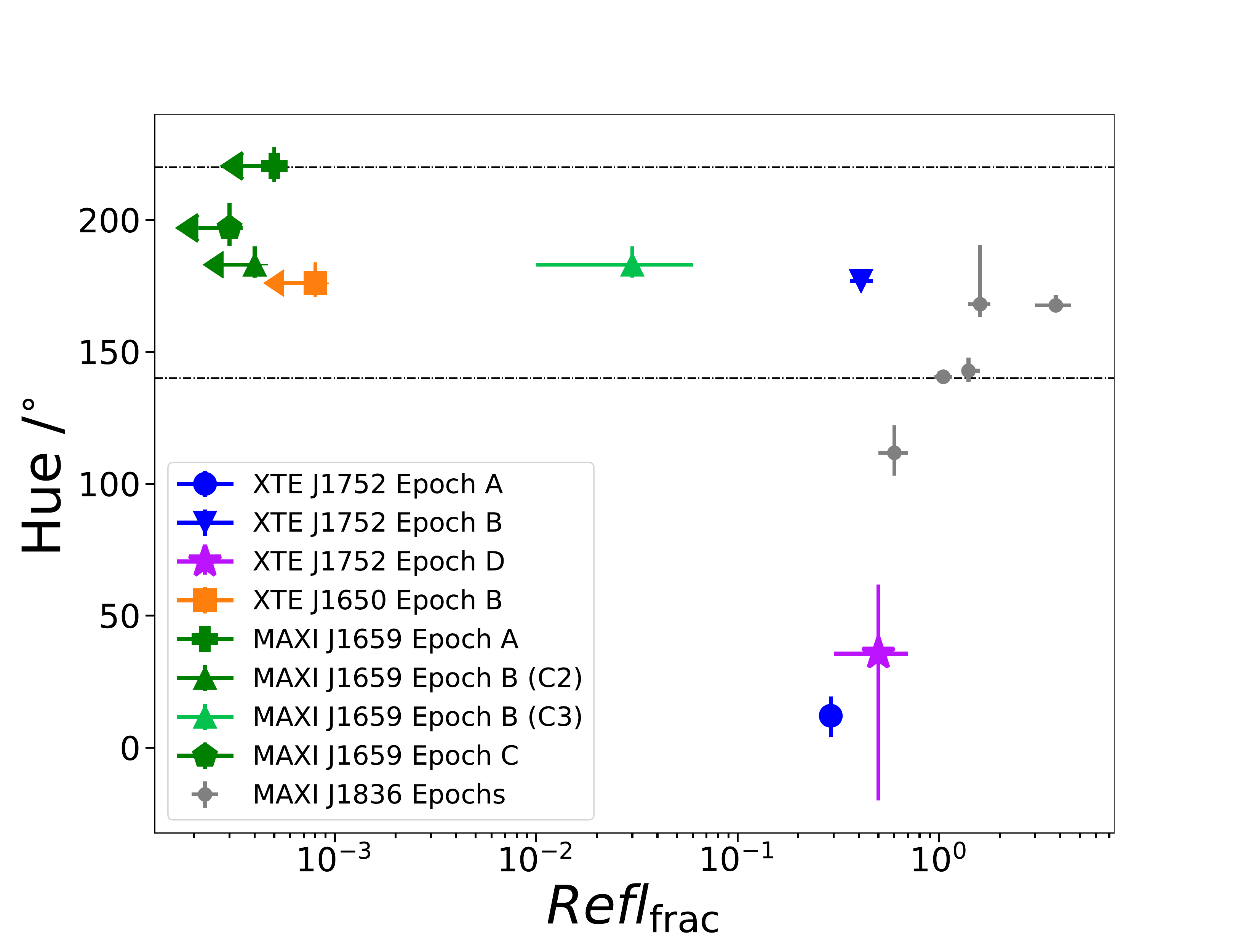}}\hfill
\caption{$\rm{Refl}_{\rm frac}$ vs. hue from the phenomenological X-ray fits (\textit{left}) and from the broadband SED joint-fits (\textit{right}). The dashed lines and the colour scheme follows Fig.~\ref{fig:r0vshue}. In both panels we also include the results from the previous study on MAXI J1836-194 (grey, taken from \citealt{lucchini2021}). Both phenomenological and physical modelling suggest a decrease in $\rm{Refl}_{\rm frac}$ as the hue increases.}\label{fig:refl}
\end{figure*}

Combining the results from the three joint-fits, we find that the hue and the coronal radius $R_0$ seem to follow a clear trend in all of the epoch clusters. Fig.~\ref{fig:r0vshue} shows a plot $R_0$ against hue values from our joint-fits. The figure also includes the results of \cite{lucchini2021}, who modelled multiple HS and HIMS SEDs of MAXI J1836-194 with the same jet model. Both studies show that larger hues generally correspond to larger $R_0$ in both HS and HIMS states. This conclusion is mainly driven by the X-ray data: a natural way of softening the X-ray Comptonisation spectrum is to lower the optical depth in the corona by increasing its size; at the same time, in the timing domain softer states usually correspond to a larger power-colour hue. One possible physical interpretation for this trend of increasing $R_0$ was proposed by \cite{lucchini2021}: $R_0$ can be thought of as a proxy for the radius of the corona-disc boundary at the jet launching region. In GRMHD simulations the size of this region is set by the pressure balance between the gas pressure at the disc/jet interface. Recent simulations show that thinner discs still launch jets, but the jets are less collimated than those launched from thicker discs \citep{liska2019bardeen}. Our findings confirm this $R_0$ vs hue trend in multiple BHXB systems and strengthen the suggestion in \citealt{lucchini2021} of a pressure-balanced coronal boundary. Furthermore, our results favour a shared geometrical configuration in the corona among BHXB sources when they evolve to similar stages during the outbursts, as also indicated by power spectral studies \citep[e.g.][]{done2006disc,done2007modelling,ingram2012modelling}. This idea about the BHXB population is demonstrated most strongly in our joint-fit for Cluster 3, in which all three epochs can be modelled with identical coronal radii, while one epoch (J1650B) is clearly more disc-dominated than the other two epochs. Our work demonstrates the potential of combining the spectral and timing analyses to characterise evolution stages.

Somewhat counter-intuitively, in MAXI J1659-152 the data can be instead fitted by a decreasing $R_0$ as the source nears the SIMS and the spectrum softens (Fig.~\ref{fig:j1659hid} - although note that in these SEDs, hue and spectral hardness are not well correlated, as shown in Fig.\ref{fig:j1659j1650hue}). In these SEDs, the joint-fit of Cluster 2 requires both the jet power $N_j$ and the electron temperature $T_e$ to drop in order to fit the data despite the decrease in $R_0$. We can speculate that this drop in $N_j$, $R_0$ and $T_e$ could be due to a gradual weakening of the compact jets, together with an increase in the disc luminosity (leading to an increased cooling of the electrons), before the state transition. This scenario is strengthened by our finding that in this cluster, the jet power $N_j$ is dropping as the hue increases. Additionally, as the jet is being quenched, the decrease of its magnetic pressure could help the coronal contraction as well. It is also possible that with the accretion rate increasing towards the Eddington limit, under the increasing radiation pressure, the disc becomes thicker again \citep[e.g.][]{abramowicz1988slim,abolmasov2015eddington,lanvcova2019puffy}, resulting again in a strongly collimated jet. 

There is a major difference in the X-ray regime between the joint-fit of Cluster 1 and the other two joint-fits: in J1752 Epoch A\&D, SSC near the jet base dominates over the scattering of disc photons. This change in the dominant radiative mechanism was also found in MAXI J1836$-$194 \citep[][]{lucchini2021}, and can be interpreted in terms of the evolution of $R_0$, which sets the number density of the radiating electrons: a more compact corona, as is the case for J1752, causes the cyclo-synchrotron radiation energy density $U_{syn}\propto L_{syn}/R^{2}_0$ to increase, which in turn results in an enhanced SSC emission. Different channels of inverse-Comptonisation can potentially cause a difference in the shape of the PSD and thus the hue. This is because the PSD, especially its low-frequency part, is closely related to the coronal response to the fluctuation of the seed spectrum (\citealt{uttley2014x}, Uttley and Malzac in prep.), and it is unlikely that the accretion rate in the disc and mass loading in the jet base would fluctuate in identical manners. A geometrical change of the accretion disc cannot solely explain the evolution of PSD during BHXB transitions \citep{ingram2011physical}, and our results point to a possible contribution from SSC inside the corona/jet base when the BHXB outburst is in the HS. Together with J1752B and other epochs during the HIMS, our results indicate a shift of the dominant Comptonisation channel should take place at some point between the HS and HIMS, due to variation in the coronal and/or truncation radii.

Another main result we find from our joint-fits is that, while large reflection fractions $\rm{Refl}_{\rm frac}$ can be found in epochs regardless of hue, low reflection fractions seem to cluster exclusively in epochs of large hue, near the HIMS/SIMS transition (Fig.~\ref{fig:refl}, although we note that this behaviour is less clear when the spectra are fitted with a phenomenological power-law, rather than our physical model). In the context of a jet model, there are two (not mutually exclusive) mechanisms that could produce this behaviour: an increase in the height of the location of the X-ray emitting region, and an increase in the bulk speed of the jet base, so that jet photons are beamed away from the disc \citep[e.g.][]{markoff2004constraining,dauser2013irradiation}. In both of these scenarios, the result would be a lowered fraction of coronal light irradiating the disc and being reprocessed in the reflection spectrum. Recently, X-ray spectral timing-analysis of the BHXB MAXI J1820+070 also suggest a similar picture by finding an increasing coronal height towards the HIMS-to-SIMS state transition \citep{wang2021disk,de2021inner} (with the caveat that a similar trend is not found in the time-averaged X-ray spectrum). Additionally, based on multi-wavelength variability, \cite{tetarenko2021measuring} also tentatively propose that the jet bulk Lorenz factor may be an increasing function of the bolometric luminosity, and \cite{wood2021varying} find evidence of an increasing ejecta speed of jet blobs as transition nears. Either mechanism is consistent with a model in which the accretion flow gets progressively more magnetically-dominated (up to the so-called Magnetically Arrested Disc, or MAD, state) as the accretion rate increases \citep[e.g.][]{tchekhovskoy2011efficient}.

The results of our joint spectral modelling have three main caveats. First, we used the phenomenological convolution model \texttt{reflect} to model the reflected spectrum from the Comptonised continuum in the jet model, rather than a more detailed, stand-alone reflection model (e.g. \texttt{relxill}, \citealt{dauser2014role}) which accounts for the reflection spectrum more accurately and self-consistently. This choice is because these advanced reflection models cannot yet be easily coupled to our jet model (\texttt{relxilllp}, for example, assumes that the emitting region is a point source, rather than an extended jet nozzle). Nevertheless, we find the trend in the reflection fraction to be in agreement with the picture derived from the detailed X-ray modelling of J1820 \citep{de2021inner,wang2021disk}, and so we consider our treatment to be a fair approximation of the reflection spectra. Second, we do not include any relativistic ray-tracing effect in our model, meaning that our constraints on the coronal geometry are subject to some systematic uncertainty. In particular, we cannot self-consistently account for variations in the coronal height or its vertical extent easily. These parameters will be explored in a future version of our model.

%%%% CONCLUSION %%%%%%%%%%%%%%%%%%%%
\section{Conclusions}
\label{sec:conclusions}

We use the steady-state, multi-zone, semi-analytical jet model \texttt{bljet} to analyse the broadband SEDs of three different BHXB sources: XTE J1752-223, MAXI J1659-152, and XTE J1650-500. We perform joint-fits of epochs with quasi-simultaneous multi-wavelength observations at HS and HIMS, clustered according to their timing power spectral colour and the associated hue values. Assuming that all of the non-thermal X-ray emission originates in the corona, associated with the jet base, we find:

\begin{itemize}[labelsep=10pt,leftmargin=.5\parindent]
    \item Multiple BHXB outbursts in different systems show a common trend in the evolution of the coronal geometry, which can be traced by the power spectral colours. During the transition from the HS to the HIMS, the corona expands due to the decrease in external pressure provided by the disc as it becomes thinner. This expansion results in a lower optical depth in the base of the jets, which in turn causes the X-ray spectrum to soften. Our results indicate a change of the dominant X-ray radiative mechanism taken place during this process, from the self-synchrotron Comptonisation to the up-scattering of disc thermal photons.
    \item We also see an indication that the corona then begins to contract close to the HIMS/SIMS transition. We propose that this contraction is related to the jet shutting down. Other possibilities include the disc becomes thicker again under the radiation pressure, re-collimating the jet base.
    \item While high reflection fractions can be found in all jetted states, from moderately bright hard states up to HIMS/SIMS transition, low reflection fractions seem to be found only near the HIMS/SIMS transition. We propose that this behaviour is caused by an increase in the vertical extent of the corona, and/or if the jet bulk speed increases towards the transition.
\end{itemize}

\section*{Acknowledgements}
We thank Dr. Phil Uttley for insightful discussions on X-ray timing analysis. This work has made use of data and software provided by: the High Energy Astrophysics Science Archive Research Center (HEASARC), which is a service of the Astrophysics Science Division at NASA/GSFC; the UK Swift Science Data Centre at the University of Leicester. This work has also made use of the Interactive Spectral Interpretation System (ISIS) maintained by Chandra X-ray Center group at MIT. M. L. and S. M. are thankful for support from an NWO (Netherlands Organisation for Scientific Research) VICI award, grant Nr. 639.043.513.

\section*{Data Availability}
All data in this paper are publicly available. The radio data were published and tabulated in \cite{corbel2004origin,brocksopp2013xte,van2013broad}. The infrared, optical, and UV data were published in \cite{curran2012disentangling,van2013broad}. The X-ray data are publicly available from HEASARC (https://heasarc.gsfc.nasa.gov/). A reproduction package is available at DOI: 10.5281/zenodo.5002124.

%%%% BIBLIOGRAPHY %%%%%%%%%%%
\bibliographystyle{mnras}
\bibliography{references}
\newpage

\section*{Appendix}
\label{sec:Appendix}
\subsection*{XTE J1752-223}

The X-ray transient XTE J1752-223 was discovered by \textit{RXTE} in 2009 October \citep{markwardt2009xte} when it went into an outburst, lasting for almost 8 months; the evolution of the outburst was typical of that of a black hole system, as shown in Fig.~\ref{fig:j1752hid}. The radio emission is consistent with a typical compact jet in the hard state and shows optically thin flares in the soft states \citep{brocksopp2013xte}. The mass of the black hole is estimated to be $9.6\pm0.8M_{\odot}$, using correlations between spectral and variability properties with GRO J1655-40 and XTE J1550-564 \citep{shaposhnikov2010discovery}, while there is a lack of dynamical mass constraint. From detailed modelling of the reflection signatures in the X-ray spectra for the month-long high-hard state with stable X-ray luminosity and hardness, the inclination angle of XTE J1752-223 is estimated to be $35^{\circ}\pm4^{\circ}$, and the galactic extinction $N_H$ to be $1.0\times10^{22}cm^{-2}$ \citep{garcia2018reflection}. 

The distance $d$ to XTE J1752-223 is not well constrained. A distance of $3.5\pm0.4$ kpcs is determined by \cite{shaposhnikov2010discovery} using the same technique for mass estimation. However, modelling of X-ray photoelectric absorption edges in the source \citep{chun2013multiwavelength} suggests a larger distance ($>$5 kpcs), and the optical detection on the companion star during the quiescence of XTE J1752-223 \citep{ratti2012black} also favours a larger distance up to 8 kpcs. A distance larger than 3.5 kpcs also agrees with the high column density $N_H$ from X-ray spectroscopy; at a distance of 3.5 kpcs on the same line of sight towards XTE J1752-223, the galactic extinction is about 60\% of what is suggested by X-ray studies \citep{chun2013multiwavelength}. Therefore, in this work we take 6 kpcs as the distance of XTE J1752-223 \citep{ratti2012black}.

\begin{table*}
\centering
%\captionsetup{justification=centering}
\caption[List of radio observations of XTE J1752-223]{List of radio observations of XTE J1752-223 analyzed in this work, detected by ATCA. Flux data in units of mJy in two frequency bands are adopted from \cite{brocksopp2013xte} and labeled according to their corresponding X-ray epochs.}
\begin{tabular}{ccccc}
\hline
Date & MJD & 5.5 GHz (mJy) & 9 GHz (mJy) & Epoch \\
\hline
2009-11-05 & 551 40.4 & $1.87\pm0.07$ & $2.05\pm0.07$ & A \\
2010-01-21 & 552 17.9 & $20.00\pm0.06$ & $21.71\pm0.04$ & B \\
2010-04-14 & 553 00.9 & $1.07\pm0.09$ & $0.09\pm0.05$ & C \\
2010-06-03 & 553 50.5 & $0.20\pm0.03$ & $0.18\pm0.03$ & D \\
\hline
\end{tabular}
\label{tb:j1752radio}
\end{table*}

\begin{table*}
\centering
\caption[List of X-ray observations of XTE J1752-223]{List of X-ray observations of XTE J1752-223 analyzed in this work, with associated epochs labeled. The UV flux of each epoch is derived from Swift\textit{UVOT} images taken during the same observations listed. On 2009-11-05, \textit{RXTE} took three successive observations and in this work we only study the first one for spectral analysis, which is the closest to the radio observation in time on the same day, but include the rest two in power-colour timing analysis. Swift has no observations coinciding Epoch B, due to the orbital restraints of the satellite.}
\begin{tabular}{cccc}
\hline
Satellite & Date & ObsID & Epoch \\
\hline
RXTE & 2009-11-05 & 94331-01-02-11 & A \\
& 2010-01-21 & 94331-01-06-02 & B \\
& 2010-04-13 & 95360-01-12-03 & C \\
& 2010-06-03 & 95702-01-07-03 & D \\
\hline
Swift & 2009-11-03 & 00031532009 & A \\
& 2010-04-14 & 00031688002 & C \\
& 2010-06-05 & 00031688022 & D \\
\hline
\end{tabular}
\label{tb:j1752xray}
\end{table*}

\textit{RXTE} monitored XTE J1752-223 throughout the entire 2009-2010 outburst. In this work we only analyze \textit{RXTE} data taken on four dates (year-month-day: 2009-11-05, 2010-01-21, 2010-04-14, 2010-06-03), highlighted by red circles in the hardness-intensity diagram (HID) in Fig.~\ref{fig:j1752hid}. These four dates correspond to epochs in which the compact jet is detected by Australia Telescope Compact Array (ATCA) \citep{brocksopp2009radio}. All the radio and X-ray observations of XTE J1752-223 used in this work are listed in Table \ref{tb:j1752radio} and Table \ref{tb:j1752xray}, and the quasi-simultaneous UV data for XTE J1752-223 is provided by Swift/\textit{UVOT}, while Swift/\textit{XRT} provides the soft X-ray spectra down to 1 keV.

\subsection*{MAXI J1659-152}

MAXI J1659-152 was detected by Swift and MAXI during its 2010 outburst \citep{mangano2010grb,negoro2010maxi}. Its orbital period of $\sim2.4$hr \citep{kuulkers2013maxi} makes it the shortest period BHXB source known \citep{kennea2010maxi}. In this work we take $6^{+1.8}_{-1.3}$ $M_{\odot}$ as the black hole mass, estimated by \cite{molla2016estimation}. The inclination angle of the system is constrained to be $65^{\circ}<i<80^{\circ}$ by using the cyclical absorption dips in X-ray lightcurves \citep{kuulkers2013maxi}. In this work we take $i=75^{\circ}$.

Like XTE J1752-223, the distance to MAXI J1659-152 is poorly constrained, ranging from 1.6 to 8.6 kpcs \citep{kennea2010maxi,miller2011x,jonker2012black,kong2012optical,kuulkers2013maxi}. Here we take it to be 6 kpcs. This value reconciles different estimations and agrees with \citealt{kong2012optical} who suggest the companion star is an M2 dwarf.

\begin{table*}
\centering
%\captionsetup{justification=centering}
\caption[List of radio observations of MAXI J1659-152]{List of radio observations of MAXI J1659-152 analyzed in this work, detected by Very Large Array (VLA). Flux data adopted from \cite{van2013broad}, labeled with Epoch A/B/C.}
\footnotesize
\begin{tabular}{ccccccc}
\hline
Date & MJD & 4.9 GHz (mJy) & 8.5 GHz (mJy)& 22 GHz (mJy)& 43 GHz (mJy)& Epoch \\
\hline
2010-09-29 & 554 68.05 & $9.88\pm0.30$ & $10.03\pm0.31$ & $11.81\pm0.71$ & $11.19\pm0.59$ & A \\
2010-10-01 & 554 70.06 & $10.29\pm0.32$ & $9.74\pm0.30$ & $8.84\pm0.49$ & $4.84\pm0.35$ & B \\
2010-10-03 & 554 61.98 & $9.23\pm0.28$ & $7.55\pm0.42$ & $7.88\pm0.42$ & $3.74\pm0.40$ & C \\
\hline
\end{tabular}
\normalsize
\label{tb:j1659radio}
\end{table*}

\begin{table*}
\centering
\caption[List of IR/UV observations of MAXI J1659-152]{List of IR/UV observations of MAXI J1659-152 used in this work. The magnitude of the data in each filter without de-reddening is adopted from \cite{van2013broad}. Data are de-reddened (See text above) before the spectral analysis.}
\begin{tabular}{ccccccc}
\hline
Date & MJD & Instrument & Filter & Magnitude & Error & Epoch \\
\hline
2010-09-29 & 55468.044 & SMARTS   & J    & 15.13  & 0.13  & A \\
& 55468.05  & SMARTS   & H    & 14.7   & 0.13  &   A \\
& 55468.494 & UVOT     & U    & 15.84  & 0.025 & A \\
\hline
2010-10-01 & 55469.79  & BOOTES-2 & R    & 16.59  & 0.06  & B \\
 & 55469.993 & SMARTS   & J    & 15.26  & 0.1   & B \\
 & 55469.999 & SMARTS   & H    & 14.88  & 0.21  & B \\
 & 55470.481 & UVOT     & UVM2 & 16.702 & 0.031 & B \\
\hline
2010-10-03 & 55471.996 & SMARTS   & J    & 15.26  & 0.16  & C \\
 & 55472.002 & SMARTS   & H    & 15.02  & 0.08  & C \\
 & 55472.12  & UVOT     & UVW1 & 16.317 & 0.027 & C \\
\hline
\end{tabular}
\label{tb:j1659ir}
\end{table*}

\begin{table*}
\centering
\caption[List of X-ray observations of MAXI J1659-152]{List of X-ray observations of MAXI J1659-152 downloaded, reduced, and analyzed in this work. From 2010-09-29 to 2010-10-15 only a fraction of the \textit{RXTE} observations has science event observation modes available for power colour analysis (see text), which is all included in this work. Timing analysis has been performed on all the listed observations in order to track the power colour evolution, while spectral analysis only considers the observations that coincide with the radio/IR/UV observations (the closest if there are multiple observations on the same day), labeled by Epoch A/B/C.}
\begin{tabular}{cccc}
\hline
Satellite & Date & ObsID & Epoch \\
\hline
RXTE & 2010-09-29 & 95358-01-02-01 & A\\
 & 2010-10-01 & 95108-01-02-00 & B\\
 & 2010-10-03 & 95108-01-05-00 & C\\
\hline
Swift & 2010-09-29 & 00434928007 & A \\
& 2010-10-01 & 00434928009 & B \\
& 2010-10-03 & 00434928011 & C \\
\hline
\end{tabular}
\label{tb:j1659xray}
\end{table*}

In \cite{van2013broad} they analyze multi-wavelength data of the entire outburst, lasting about 40 days. These authors estimates a line-of-sight column density $N_H=(0.319\pm0.009)\times10^{22}$ cm$^{-2}$. The multi-wavelength data are presented in Table \ref{tb:j1659radio} and Table \ref{tb:j1659ir}. The three epochs in which radio, IR/optical/UV and X-ray data are available and our jet model is applicable, are selected and highlighted by red circles in Fig.~\ref{fig:j1659hid}. The X-ray observations used in the spectral analysis by this work are listed in Table \ref{tb:j1659xray}). Within the selected time interval, the jet break frequency moved from IR to radio band \citep{van2013broad}, which can be an indication of the jet particle acceleration region extending further out from the black hole (\citealt{lucchini2021}, also see the Model section). 

\subsection*{XTE J1650-500}

XTE J1650-500 was first discovered by \textit{RXTE} during its 2001 outburst \citep{remillard2001xte}. We adopt 5.1 $M_{\odot}$ for the mass of the black hole, as estimated by \cite{slany2008mass} and within the mass limit estimated by \cite{orosz2004orbital} ($2.7<M_{BH}<7.3$ $M_{\odot}$). We take an intermediate inclination $i$=$45^{\circ}$, following estimates from modelling its X-ray reflection spectra \citep{miller2002evidence,miniutti2004relativistic}. The distance of XTE J1650-500 is estimated to be $2.6\pm0.7$ kpcs by empirically studying the X-ray luminosity of BHXBs during the state transitions \citep{homan2006xmm}. $N_H$ estimations vary among studies but all favour a moderate absorption \citep[e.g.][]{miller2002evidence,montanari2009bepposax}; here we take $N_H=(0.5\pm0.1)\times10^{22}cm^{-2}$ from \citep{miniutti2004relativistic}.

\begin{table*}
\centering
%\captionsetup{justification=centering}
\caption[List of radio observations of XTE J1650-500]{List of radio observations of XTE J1650-500 used in this work. All flux density data is in units of mJy, adopted from \cite{corbel2004origin} and labeled by Epoch A/B.}
\begin{tabular}{ccccccc}
\hline
Date & MJD & 1384 MHz & 2496 MHz & 4800 MHz & 8640 MHz & Epoch \\
\hline
2001-09-08 & 521 60.81 & $4.08\pm0.20$ & $5.30\pm0.15$ & $5.28\pm0.10$ & $4.48\pm0.10$ & A \\
2001-09-24 & 521 77.01 & - & - & $0.83\pm0.10$ & $0.77\pm0.10$ & B \\
\hline
\end{tabular}
\label{tb:j1650radio}
\end{table*}

\begin{table*}
\centering
\caption[List of IR observations of XTE J1650-500]{List of IR observations of XTE J1650-500 used in this work. The data magnitudes before de-reddening are adopted from \cite{curran2012disentangling}. The data are de-reddened before the spectral analysis.}
\begin{tabular}{cccccc}
\hline
Date & MJD & Filter & Magnitude & Error & Epoch \\
2001-09-08 & 521 61.02 & J       & 14.42 & 0.11 & A \\
 & 521 61.03 & H       & 13.79 & 0.11 &  A\\
 & 521 61.04 & $K_{S}$ & 13.29 & 0.13 &  A\\
\hline
2001-09-25 & 521 76.99 & J       & 14.65 & 0.09 & B \\
 & 521 76.99 & H       & 14.16 & 0.12 &  B\\
 & 521 77.00 & $K_{S}$ & 13.82 & 0.12 &  B\\

\hline
\end{tabular}
\label{tb:j1650ir}
\end{table*}

\begin{table*}
\centering
\caption[List of X-ray observations of XTE J1650-500]{List of X-ray observations of XTE J1650-500 downloaded, reduced and analyzed in this work. For this source in X-ray, only \textit{RXTE} data were analyzed. Timing analysis has been performed on all the listed observations in order to track the power colour evolution, while spectral analysis only considers the observations that coincide with the radio/IR/UV observations, labeled by Epoch A/B.}
\begin{tabular}{cccc}
\hline
Date & ObsID & Epoch \\
\hline
2001-09-08 & 60113-01-03-00 &  A \\
2001-09-25 & 60113-01-19-00 &  B \\
\hline
\end{tabular}
\label{tb:j1650xray}
\end{table*}

Fig.~\ref{fig:j1650hid} shows the path of the source on the HID during its outburst, using \textit{RXTE} data. In total there are eight radio observations simultaneous with X-ray observations, but only two epochs have simultaneous IR data at the HS or HIMS, indicated by red circles in Fig.~\ref{fig:j1650hid}. Note that we only analyze \textit{RXTE} observations of these two epochs mentioned above. Multi-wavelength spectral analysis (radio/O/IR/X-ray) is only performed on the two highlighted epochs. See Table \ref{tb:j1650radio}, Table \ref{tb:j1650ir}, and Table \ref{tb:j1650xray} for the lists of radio, IR, and X-ray observations used in this work respectively.

\begin{table*}
\renewcommand{\arraystretch}{1.5}
\centering
\caption[111]{Best-fitting parameters of clusters considered in this work. Similar settings as Table \ref{tb:mwl}, but with $R_0$ and $R_{\rm in}$ un-tied among all the epochs.}
\resizebox{\textwidth}{!}{%
\begin{tabular}{cc|cc|ccc|ccc}
\hline
&&\multicolumn{2}{c|}{Cluster 1}&\multicolumn{3}{c|}{Cluster 2}&\multicolumn{3}{c}{Cluster 3}\\
Model Component&Parameter&J1752 A&J1752 D&J1659 A&J1659 B&J1659 C&J1752 B&J1659 B&J1650 B\\
\hline
TBabs&$N_H$ ($10^{22}cm^{-2}$)&\multicolumn{2}{c|}{(1.0)}&\multicolumn{3}{c|}{(0.32)}&(1.0)&(0.32)&(0.5)\\
\hline
Bljet%&$M_{BH}$ ($M_{\odot}$)&\multicolumn{2}{c|}{(9.6)}&\multicolumn{3}{c|}{(6)}&(9.6)&(6)&(5.1)\\
%&$Incl$ ($^{\circ}$)&\multicolumn{2}{c|}{(35)}&\multicolumn{3}{c|}{(75)}&(35)&(75)&(45)\\
%&$d$ ($kpc$)&\multicolumn{2}{c|}{(6)}&\multicolumn{3}{c|}{(6)}&(6)&(6)&(2.6)\\
&$f_{\rm pl}$&$8.6^{+0.2}_{-0.3}$&$15^{+1}_{-2}$&\multicolumn{3}{c|}{(0)}&$2.95^{+0.08}_{-0.27}$&(0)&$3.3^{+0.4}_{-0.2}$\\
&$N_j$&$0.024^{+0.001}_{-0.002}$&$0.012^{+0.003}_{-0.003}$&$0.0393^{+0.0008}_{-0.0006}$&$0.033^{+0.001}_{-0.001}$&$0.0294^{+0.0006}_{-0.0005}$&$0.038^{+0.001}_{-0.003}$&$0.0334^{+0.0006}_{-0.0005}$&$0.0032^{+0.0002}_{-0.0002}$\\
&$R_0$ ($R_g$)&$12.5^{+0.5}_{-0.7}$&$10^{+2}_{-2}$&$40.2^{+0.6}_{-0.7}$&$34.4^{+0.5}_{-0.5}$&$31.1^{+0.6}_{-0.5}$&$35^{+1}_{-2}$&$34.5^{+0.5}_{-0.6}$&$35^{+2}_{-1}$\\
&$T_e$ ($keV$)&$244^{+13}_{-7}$&$187^{+27}_{-24}$&$160^{+4}_{-3}$&$104^{+4}_{-3}$&$116^{+3}_{-4}$&$45^{+2}_{-3}$&$100^{+3}_{-5}$&$319^{+7}_{-7}$\\
&$R_{\rm in}$ ($R_g$)&$182^{+12}_{-19}$&$122^{+14}_{-18}$&$19.5^{+0.3}_{-0.3}$&$24.2^{+0.4}_{-0.4}$&$15.2^{+0.4}_{-0.3}$&$10.2^{+0.5}_{-0.4}$&$24.0^{+0.4}_{-0.4}$&$12.3^{+0.2}_{-0.1}$\\
&$L_{disc}$ ($L_{Edd}$)&$0.014^{+0.003}_{-0.002}$&$0.005^{+0.001}_{-0.001}$&$0.087^{+0.001}_{-0.001}$&$0.112^{+0.002}_{-0.001}$&$0.105^{+0.002}_{-0.003}$&$0.070^{+0.001}_{-0.002}$&$0.112^{+0.003}_{-0.001}$&$0.0546^{+0.0007}_{-0.0006}$\\
\hline
reflect&$\rm{Refl}_{\rm frac}$&$0.31^{+0.02}_{-0.02}$&$0.4^{+0.3}_{-0.2}$&$4.4^{+8.0}_{-3.5}\times10^{-3}$&$0.03^{+0.02}_{-0.02}$&$0.018^{+0.035}_{-0.014}$&$0.42^{+0.06}_{-0.05}$&$0.03^{+0.03}_{-0.02}$&$0.02^{+0.06}_{-0.01}$\\
\hline
gaussian&$E_{center}$ (keV)&\multicolumn{2}{c|}{(6.4)}&\multicolumn{3}{c|}{(6.4)}&\multicolumn{3}{c}{(6.4)}\\
&$\sigma_{gauss}$&$0.23^{+0.29}_{-0.16}$&$0.006^{+0.009}_{-0.004}$&$0.8^{+0.2}_{-0.1}$&$0.2^{+0.1}_{-0.1}$&$1.1^{+0.3}_{-0.3}$&$0.86^{+0.04}_{-0.05}$&$<0.01$&$1.10^{+0.03}_{-0.02}$\\
&$A_{gauss}$&$7^{+3}_{-3}\times10^{-4}$&$1.8^{+1.0}_{-0.8}\times10^{-4}$&$3.5^{+0.6}_{-0.6}\times10^{-3}$&$1.4^{+0.3}_{-0.3}\times10^{-3}$&$5^{+2}_{-1}\times10^{-3}$&$18.5^{+1.0}_{-1.2}\times10^{-3}$&$1.2^{+0.3}_{-0.2}\times10^{-3}$&$2.00^{+0.05}_{-0.04}\times10^{-2}$\\
\hline
Black Body&$T_{bbody}$ ($K$)&$4.6^{+1.0}_{-0.8}\times10^{3}$&-&\multicolumn{3}{c|}{(30000)}&-&(30000)&-\\
&$L_{bbody}$ ($erg/s$)&$2.4^{+1.4}_{-0.6}\times10^{36}$&-&\multicolumn{3}{c|}{$1.02^{+0.04}_{-0.05}\times10^{36}$}&-&$1.19^{+0.07}_{-0.06}\times10^{36}$&-\\
\hline 
$\chi^2$/d.o.f&&\multicolumn{2}{c|}{402/238}&\multicolumn{3}{c|}{1770/704}&\multicolumn{3}{c}{919/335}\\
\hline
\end{tabular}}
\label{tb:mwl_a}
\end{table*}

\end{document}